\documentclass[longauth,useAMS,usenatbib]{mn2e}
\usepackage{graphicx,bm,setspace,amssymb,color,amsmath,url}
\numberwithin{equation}{section}

\newcommand{\mv}[1]{{\bm #1}}
\newcommand{\ii}{\mathrm{i}}

\newcommand{\me}[1]{\exp \left[ {#1} \right]}

\newcommand{\ips}[1]{\left|\widetilde{\phi}\left({#1}\right)\right|^2}

\newcommand{\lr}[1]{\left\langle #1 \right\rangle}

\newcommand{\bcdot}{\bm{\cdot}}

\title[Scintillation noise in 21-cm observations]
  {Scintillation noise power spectrum and its impact on high redshift 21-cm observations} 

\author[Vedantham \& Koopmans]
{H.K.~Vedantham$^{a,b}$\thanks{E-mail: harish@astro.caltech.edu} and L.V.E.~Koopmans$^{a}$\\
$^{a}$Kapteyn Astronomical Institute, University of Groningen, P.O. Box 800, 9700 AV Groningen, The Netherlands\\
$^{b}$California Institute of Technology, 1200 East California Boulevard, Pasadena, California 91125, USA}

\begin{document}
%
\date{\today}
\pagerange{\pageref{firstpage}--\pageref{lastpage}} \pubyear{2014}
\def\LaTeX{L\kern-.36em\raise.3ex\hbox{a}\kern-.15em
    T\kern-.1667em\lower.7ex\hbox{E}\kern-.125emX}
\newtheorem{theorem}{Theorem}[section]
\label{firstpage}
\maketitle
%
%
%
%
%
\begin{abstract}
Visibility scintillation resulting from wave propagation through the turbulent ionosphere can be an important sources of noise at low radio frequencies ($\nu\lesssim 200$~MHz). Many low frequency experiments are underway to detect the power spectrum of brightness temperature fluctuations of the neutral-hydrogen $21$-cm signal from the Epoch of Reionization (EOR: $12\gtrsim z\gtrsim 7$, $100\lesssim \nu \lesssim 175$~MHz). In this paper, we derive scintillation noise power-spectra in such experiments while taking into account the effects of typical data processing operations such as self-calibration and Fourier synthesis. We find that for minimally redundant arrays such as LOFAR and MWA, scintillation noise is of the same order of magnitude as thermal noise, has a spectral coherence dictated by stretching of the snapshot $uv$-coverage with frequency, and thus is confined to the well known wedge-like structure in the cylindrical ($2$-dimensional) power spectrum space. Compact, fully redundant ($d_{\rm core}\lesssim r_{\rm F} \approx 300$~m at $150$~MHz) arrays such as HERA and SKA-LOW (core) will be scintillation noise dominated at all baselines, but the spatial and frequency coherence of this noise will allow it to be removed along with spectrally smooth foregrounds.
\end{abstract}

%
%
%
%
\section{Introduction}
\label{ch4sec:intro}
Observations of the highly redshifted $21$-cm signal from the epochs of Cosmic Dawn (CD; $35\gtrsim z \gtrsim 15$) and Reionization (EoR; $15\gtrsim z \gtrsim 6$) are expected to revolutionise our understanding of structure formation in the first billion years of the Universe's history \citep{furlanetto2006}. To achieve this, observations with radio telescopes such as LOFAR \citep{lofar}, PAPER \citep{paper}, MWA \citep{mwa}, PAST \citep{past2004}, and GMRT \citep{paciga2013} are ongoing, while a next generation of larger telescopes such as NenuFar \citep{nenufar}, HERA\footnote{\url{http://reionization.org}} \citep{hera} and SKA\footnote{\url{http://www.skatelescope.org}} are being planned or commissioned. Ongoing CD and EoR experiments aim to constrain the brightness-temperature fluctuations in the $21$-cm signal statistically by measuring its two-point correlation function, or equivalently the angular power-spectrum. Even a statistical measurement will require several hundreds to thousands (array dependent) of hours of integration time owing to the faintness of the predicted signal as compared to astrophysical foreground emission, prompting a rigorous analysis of various sources of noise and systematic biases.\\

The $21$-cm signal from CD and EoR epochs is redshifted to low radio frequencies ($40 \lesssim \nu \lesssim 200$~MHz) where ionospheric propagation effects are important. Recently, \citet{speckle} showed that propagation through a turbulent ionosphere results in stochasticity, or uncertainty, in interferometric visibilities, and that this additional `scintillation noise' can, under reasonable observing conditions, be larger than thermal uncertainties at low radio frequencies ($\nu<200$~MHz). The principal aim of this paper is to apply the analytical results of \citet{speckle} to the case of high redshift $21$-cm observations, and to forecast the scintillation noise bias in $21$-cm power spectra.\\

Due to the large number of equations and associated variables used in this paper, we have summarised our main results in Section \ref{ch4sec:mainresults}, and also listed the variables and their meanings at the end of the paper for easy reference. The rest of the paper is organised as follows. Section \ref{ch4sec:basics} summarises the basic analytical expressions to compute the statistics of visibility scintillation. These expressions were derived in \citet{speckle} for fields of view (FOV) of about $10$~degrees at meter wavelengths. In Section \ref{ch4sec:wfe} we extent these results to an arbitrarily large FOV. In Section \ref{ch4sec:fse}, we describe the effects of Fourier synthesis (time/frequency averaging and gridding) on scintillation noise. In addition to Fourier synthesis, interferometric arrays also employ self-calibration to alleviate ionospheric and instrumental corruptions. We discuss the calibratability of scintillation noise and the associated implications in Section \ref{ch4sec:cal}. In Section \ref{ch4sec:snps}, we use the results of all the preceding sections to make scintillation noise forecasts in the cosmological wavenumber space in which the $21$-cm power spectrum will eventually be determined. Finally in Section \ref{ch4sec:concl}, we draw conclusions and recommendations for future work.
%
%
%
%
%
%
\section{Basic results}
\label{ch4sec:basics}
In this section, we summarise the equations describing the statistics of ionospheric phase fluctuations and the resulting visibility scintillation. The visibility statistics are computed by employing the Fresnel-Huygens principle, where each point on the ionospheric phase screen is considered to be a secondary radiator of spherical waves. The ensuing waves are all coherently summed up at the observer's location to compute the emergent field using Kirchhoff's diffraction integral \citep{bornwolf}. To keep the analytical derivations tractable, we Taylor expanded the path-length between the secondary radiators and the observer to quadratic order, which corresponds to the case of Fresnel diffraction \citep{bornwolf}. Higher order terms in the path-length become comparable to a wavelength, if the field of view (FOV) exceeds about $10$~degrees for meter-wavelengths. We generalise the results from this section to an arbitrary field of view in Section \ref{ch4sec:wfe}.
\subsection{Visibility scintillation and coherence}
We model the additional electromagnetic phase introduced by the ionospheric as a $2$-dimensional (thin-screen approximation) Gaussian random field, with a total variance $\phi_0^2$, and an isotropic power spectrum of spatial fluctuations given by
\begin{equation}
\label{ch4eqn:vonkarman}
\ips{k} = \frac{5\phi_0^2 }{6\upi k_{\rm o}^2} \left[\left( \frac{k}{k_{\rm o}} \right)^2 +1 \right]^{-11/6},
\end{equation}
where $k=1/L$ is the length of the $2$-dimensional wavenumber vector $\mv{k}$ for a wavelength of $L$, $k_{\rm o}$ is the outer scale, or energy injection scale, for the turbulence. For $k \gtrsim k_{\rm o}$ the power spectrum follows the usual Kolmogorov $11/3$-index power law. The outer scale is not uniquely determined but is typically much larger than the other relevant length scales in our calculations (Fresnel-length $r_{\rm F}$ and baseline length $b$), and thus does not significantly influence the results. We will choose it to be $100$~km in this paper\footnote{Measurements of the phase structure function at 150~MHz with LOFAR affirm this assumption (Maaijke Mevius, priv. comm.).}. Though the power spectrum is then completely defined by the phase variance $\phi_0^2$, a related quantity called the diffractive scale is easier to measure. The variance of phase difference between $2$ points separated by the diffractive scale is defined to be $1$~radian~squared. In that case, $\phi_0^2$ and $r_{\rm diff}$ are related by
\begin{equation}
r_{\rm diff} = \frac{1}{\upi k_{\rm o}}\left( \frac{\Gamma(11/6)}{2\Gamma(1/6)\phi_0^2}\right)^{3/5},
\end{equation}
where $\Gamma(.)$ is the Gamma function.\\

Let the visibility of a source in direction $\mv{l}$ measured on baseline\footnote{We use the term `baseline' to denote the physical separation between a given pair of antennas. Hence `redundant baselines' are considered to be separate baselines in this definition.}  $\mv{b}$ be $V_{\rm M}(\mv{b},\mv{l})$. If the visibility in the absence of any ionospheric effects is $V_{\rm T}(\mv{b},\mv{l})$, then we can show the following results for the statistics of $V_{\rm M}(\mv{b},\mv{l})$ \citep[see e.g. ][]{speckle}:
\begin{equation}
\label{ch4eqn:expect}
\lr{V_{\rm M}(\mv{b},\mv{l})} = V_{\rm T}(\mv{b},\mv{l})\me{-\frac{1}{2}\mathcal{D}(b)},
\end{equation}
where the expectation is taken over an ensemble of ionospheric phase-screen realisations, and $\mathcal{D}(b)$ is the ionospheric phase structure-function on a baseline of length $b$ which may be approximated for $\upi k_{\rm o} r \ll 1$ as
\begin{equation}
\mathcal{D}(r) = \left( \frac{r}{r_{\rm diff}}\right)^{5/3}.
\end{equation}
The covariance function of $V_{\rm M}$ on angular, spatial, and temporal dimensions under conditions of weak refractive scintillation is \citep{speckle}
\begin{eqnarray}
\label{ch4eqn:general_cov}
{\rm Cov}[V_{\rm M},\Delta\mv{s}] &=& 4\int {\rm d}^2\mv{q}\ips{\mv{q}}\sin^2\left(\upi\lambda h \mv{q}^2-\upi\mv{b}\bcdot\mv{q}\right) \nonumber \\
&& \times \,  P_k(\mv{b},\Delta\mv{s}) \me{-\ii 2\upi \mv{q}\bcdot \Delta\mv{s}},
\end{eqnarray}
where $h$ is the height of the ionospheric screen, $P_k(\mv{b},\Delta\mv{s})$ is the sky-brightness power spectrum (defined below), and the vector $\Delta\mv{s}$ can be interpreted as any of the following:
\begin{enumerate}
\item The spatial separation between two baselines of the same length and orientation, thus yielding the spatial covariance of visibility scintillation, for which $P_k(\mv{b},\Delta\mv{s}) = V_{\rm T}(\mv{b})V^\ast_{\rm T}(\mv{b'})=\left|V_{\rm T}(\mv{b}) \right|^2$ where $\Delta\mv{s}$ is the displacement between the two redundant baselines.\\
\item  $\Delta\mv{s}=\mv{v}\tau$, where $\mv{v}$ is the bulk-velocity with which the ionospheric turbulence moves, thus yielding the temporal coherence of visibility scintillation on a time-scale $\tau$, for which $P_k(\mv{b},\Delta\mv{s}) = V_{\rm T}(\mv{b},t=0)V^\ast_{\rm T}(\mv{b},t=\tau)$. Here we are assuming that the ionospheric turbulence does not evolve significantly during the time it crosses the interferometer array, which is the widely used Taylor's `frozen irregularities' assumption \citep{taylor38}. \\
\item  $\Delta\mv{s}=h\Delta\mv{l}$ where $\Delta\mv{l}$ is the angular separation of any two sources, thus yielding the angular coherence of visibility scintillation, for which $P_k(\mv{b},\Delta\mv{s}) = V_{\rm T}(\mv{b},\mv{l})V^\ast_{\rm T}(\mv{b},\mv{l'})$, and $\Delta\mv{l}=\mv{l}-\mv{l'}$.\\
\end{enumerate}
Equation (\ref{ch4eqn:expect}) and (\ref{ch4eqn:general_cov}) have been derived using Fresnel's approximation to Kirchhoff's diffraction integrals \citep{bornwolf}, and include the effects of both amplitude and phase scintillations. The natural `coherence scale' in the emergent field is given by the Fresnel length $r_{\rm F}=\sqrt{\frac{\lambda h}{2\pi}}$. Numerically evaluating the Fourier-transform in equation (\ref{ch4eqn:general_cov}), gives the following result: If $b\lesssim r_{\rm F}$, then the ${\rm Cov}[V_{\rm M},\Delta\mv{s}]$ reaches half of its peak value (attained at $\mv{s}=\mv{0}$) for $s\approx r_{\rm F}$, and for $b\gtrsim r_{\rm F}$, ${\rm Cov}[V_{\rm M},\Delta\mv{s}]$ reaches half of its peak value for $s\approx 2b$ if $\mv{s}$ is parallel to $\mv{b}$ and for $s\approx b$ is $\mv{s}$ is perpendicular to $\mv{b}$.\\

Additionally, one can integrate the angular coherence function defined above over $\Delta\mv{l}$ to obtain the scintillation noise variance from the entire sky:
\begin{eqnarray}
\label{ch4eqn:scint_fullps}
\sigma^2[V_{\rm M}(\mv{b})] &=& 4\int {\rm d}^2\mv{q} \ips{\mv{q}} \sin^2(\upi\lambda h \mv{q}^2-\upi\mv{b}\bcdot\mv{q}) \nonumber \\
&& \times \, \left|V_{\rm T}(\mv{b}-\lambda h \mv{q} )\right|^2
\end{eqnarray}

where $\left| V_{\rm T}(.) \right| ^2$ is the power spectrum of the sky.\\

The integrated brightness temperature of point-like sources at $150$~MHz is about $18$~K \citep{vernstrom2011} which is substantially larger than the brightness of diffuse Galactic emission on baselines of interest here ($b\gtrsim 10\lambda$). We thus assume that the sky is composed of Poisson distributed point-like sources which makes the sky power spectrum independent of the baseline length. It can then be brought out of the integral in equation (\ref{ch4eqn:scint_fullps}) as $S^2_{\rm eff} \equiv \left|V_{\rm T}(\mv{b}-\lambda h \mv{q} )\right|^2$, where $S_{\rm eff}$ depends on the flux-density distribution of sources and may be called the effective scintillating flux for a given random source ensemble. Hence, we get
\begin{eqnarray}
\label{ch4eqn:scint_seff}
\sigma^2[V_{\rm M}(\mv{b})] &\equiv& S^2_{\rm eff}\sigma^2_{\rm fr}[\mv{b}] \nonumber \\
\sigma^2_{\rm fr}[\mv{b}]&=&4\int {\rm d}^2\mv{q} \ips{\mv{q}} \sin^2(\upi\lambda h \mv{q}^2-\upi\mv{b}\bcdot\mv{q})
\end{eqnarray}
where $\sigma^2_{\rm fr}[\mv{b}]$ is the fractional scintillation variance which may be interpreted as the visibility variance due to scintillation of a $S=1$~Jy source. A useful approximation that is valid for long baselines is\footnote{The maximum bound on $\sigma^2_{\rm fr}[\mv{b}]$ under this approximation is $2\phi_0^2$.}
\begin{equation}
\sigma^2_{\rm fr}[\mv{b}] \approx \left( \frac{b}{r_{\rm diff}}\right)^{5/3}\textrm{  for } b\gg r_{\rm F}
\end{equation}
The above results were derived in \citet{speckle} for fields of view of $\lesssim 10$~degrees at metre-wavelengths, but as shown in section \ref{ch4sec:wfe}, they are also good approximations for larger fields of view.
\subsection{The effective scintillating flux}
To compute realistic values of scintillation noise, we will use the differential source counts given by \citep[see][]{speckle}
\begin{equation}
\label{ch4eqn:dsc}
\frac{{\rm d}^2N(S)}{{\rm d}S{\rm d}\Omega} =C \nu^{-\beta} S^{-\alpha}.
\end{equation}
Choosing appropriate values for the constants $C$, $\beta$, and $\alpha$ based on the differential source counts of \citet{windhorst1985}, and low frequency spectral indices measured by \citep{vlss},  we use the following source-counts in this paper
\begin{equation}
\label{ch4eqn:dscvals}
\frac{{\rm d}^2N(S)}{{\rm d}S{\rm d}\Omega} \approx 3\times 10^3\,\left( \frac{\nu}{150\textrm{ MHz}}\right)^{-0.8} \left( \frac{S}{1\textrm{ Jy}} \right)^{-2.5} \textrm{ Jy$^{-1}$sr$^{-1}$}.
\end{equation}
If the primary beam at frequency $\nu$, and direction $\mv{l}$ for a circular primary aperture of diameter $d_{\rm prim}$ is $B(d_{\rm prim},\nu,\mv{l})$, then the source counts of equation (\ref{ch4eqn:dsc}) can be converted into the number of sources with apparent flux in the range $S$ to $S+{\rm d}S$ from the entire sky as
\begin{equation}
\label{ch4eqn:dnds_appflux}
\frac{{\rm d}N(S)}{{\rm d}S} = C\nu^{-\beta}B_{\rm eff}(d_{\rm prim},\nu)S^{-\alpha},
\end{equation}
where $B_{\rm eff}$ is the effective beam for scintillation noise calculations, and is given by
\begin{equation}
\label{ch4eqn:beff}
B_{\rm eff}(d_{\rm prim},\nu) = \int\int_{2\upi}	{\rm d}\Omega B^{\alpha-1}(d_{\rm prim},\nu,\mv{l}), 
\end{equation}
where ${\rm d}\Omega$ is the differential solid angle. The above equation assumes that the source counts from equation (\ref{ch4eqn:dsc}) hold for all values of the flux density $S$. Though the source counts must cut off at some small flux-density value such that the integrated flux is bounded, our assumption is inconsequential since as we shall soon see, scintillation is by dominated the brighter sources. For simplicity, we split the source population into two parts: (i) bright sources ($S>S_{\rm max}$) which are part of a sky model and whose scintillation noise has been calibrated out, and (ii) weaker sources ($S<S_{\rm max}$) whose aggregate scintillation noise will remain in the data post calibration. The effective scintillating flux due to all the weaker uncalibrated sources becomes
\begin{equation}
\label{ch4eqn:seff}
S^2_{\rm eff}(d_{\rm prim},\nu) = \frac{CB_{\rm eff}(d_{\rm prim},\nu) \nu^{-\beta}}{3-\alpha} S_{\rm max}^{3-\alpha}(d_{\rm prim},\nu).
\end{equation}
The value of $S_{\rm max}$ largely depends on the thermal noise which at  low frequencies is typically dominated by sky noise (as opposed to receiver noise). We assume a sky temperature of
\begin{equation}
\label{ch4eqn:tsky}
T_{\rm sky}(\nu) = T_0 \nu^{-\gamma}\,\,\, {\rm K}
\end{equation}
where $T_0$ and $\gamma$ are constants. For calculations in this paper, we will choose these constants to be \citep{landecker1970}
\begin{equation}
\label{ch4eqn:tskyvals}
T_{\rm sky}(\nu) \approx 300\left( \frac{\nu}{150\textrm{ MHz}}\right)^{-2.5}\,\,\,\ {\rm K,}
\end{equation}
For a fully filled primary aperture of diameter $d_{\rm prim}$, a sky brightness temperature of $T_{\rm sky}$ gives a system equivalent flux density of 
\begin{equation}
\label{ch4eqn:sefd}
{\rm SEFD}(\nu) = \frac{2kT_{\rm sky}(\nu)}{\upi d^2_{\rm prim}/4},
\end{equation}
where $k$ is Boltzmann's constant. On using $T_{\rm sky}(\nu)$ from equation \ref{ch4eqn:tskyvals}, we get
\begin{equation}
\label{ch4eqn:sefdvals}
{\rm SEFD}(\nu) \approx 1.2 \left( \frac{d_{\rm prim}}{30\textrm{ m}}\right)^{-2}\left( \frac{\nu}{150\textrm{ MHz}}\right)^{-2.5}\,\,\,{\rm kJy.}
\end{equation}

Assuming that scintillation noise from all sources that present a signal to noise ratio of $\zeta$ or higher per visibility is perfectly removed using self-calibration, $S_{\rm max}$ can be written as
\begin{equation}
\label{ch4eqn:smax}
S_{\rm max}(d_{\rm prim},\nu) = \zeta\frac{{\rm SEFD}}{\sqrt{2\Delta\nu\Delta\tau}},
\end{equation}
where $\Delta\nu$ and $\Delta\tau$ are the frequency and time cadence for calibration solutions. If we choose $\zeta=5$, $\Delta\nu=1$~MHz (typical channel-width for self-calibration) and $\Delta\tau=2$~sec (typical scintillation decorrelation timescale for short baselines), then $S_{\rm max}$ can be written as
\begin{equation}
\label{ch4eqn:smax_vals}
S_{\rm max}(d_{\rm prim},\nu) \approx 3\left( \frac{d_{\rm prim}}{30\textrm{ m}}\right)^{-2} \left( \frac{\nu}{150\textrm{ MHz}}\right)^{-2.5}\,\textrm{ Jy.}
\end{equation}
Finally, using this value for $S_{\rm max}$ in equation (\ref{ch4eqn:seff}), we can write the effective scintillating flux as
\begin{equation}
\label{ch4eqn:seffvals}
S_{\rm eff} \approx 5.86 \left( \frac{d_{\rm prim}}{30\textrm{ meter}}\right)^{-1.5}\left( \frac{\nu}{150\textrm{ MHz}}\right)^{-2.025}\,\,\, {\rm Jy.}
\end{equation} 
We note here that while thermal noise per visibility scales with aperture size as $d^{-2}_{\rm prim}$, the effective scintillating flux scales as $d^{-1.5}_{\rm prim}$. This comes about since decreasing $d_{\rm prim}$, increases the number of sources contributing to scintillation noise (increased beam-width) whose rms flux scales as $d^{-1}_{\rm prim}$. In addition, decreasing $d_{\rm prim}$ also increases the thermal noise per visibility, which increases $S_{\rm max}$ which results in an additional scaling dependence of $d^{-0.5}_{\rm prim}$.
%
%
%
%
%
\section{Widefield effects}
\label{ch4sec:wfe}
The results in the preceding section are accurate for fields of views of about $10$~deg at meter-wavelengths. While LOFAR's high band antenna stations (HBA) are within this limit, other arrays such as the MWA, PAST, and PAPER have FOVs that exceed this limit. In addition to this, we have not yet incorporated the effects of an increased distance to the ionospheric screen, and an increased propagation path-length through the ionosphere\footnote{Though were are working with the thin-screen approximation, we must still include the effects increased scattering due to a larger path length through the ionosphere.} for off-zenith sources. Finally, we assumed a plane-parallel diffraction screen in our derivations, which is violated in wider FOV cases due to the curvature of the Earth's ionosphere. While inclusion of all this effects in closed form may be analytically intractable, in this section we extend equation (\ref{ch4eqn:general_cov}) to the generic all-sky field of view case by making certain justified simplifications.
\subsection{A slant geometry}
\begin{figure}
\centering
\includegraphics[width=0.5\linewidth]{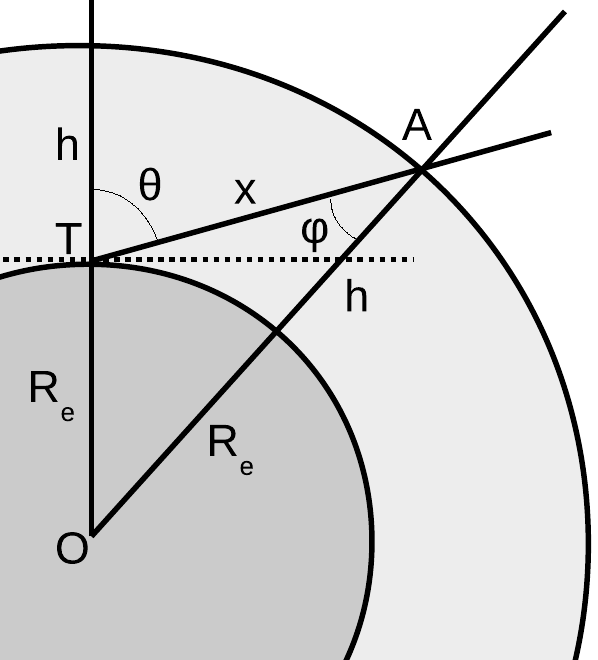}
\caption{A depiction of an off-nadir viewing geometry for scintillation noise calculations. A zenith angle ($\theta$) increase results in an increase in the distance to the scattering screen $h(\theta)$ and hence an increase in the Fresnel scale $r_{\rm F}(\theta)=\sqrt{\lambda h(\theta)/(2\upi)}$. Increasing $\theta$ also results in an increase in the path-length through the ionospheric turbulence and in effect, scales the ionospheric phase-power spectrum by $\sec(\theta)$.\label{ch4fig:slant_geom}}
\end{figure}

To understand wide-field effects, we consider a slant viewing geometry as shown in Fig. \ref{ch4fig:slant_geom} as opposed to a zenith viewing case. Because the real ionosphere is not a thin-screen but has a finite thickness, the propagation length approximately follows a $\sec\theta$ dependence, though this approximation is violated for large zenith angles. We may incorporate this effect by scaling the ionospheric power spectrum by a factor of $\sec\theta$. In addition, the associated increase in Fresnel length ensuing from a $\sec\theta$ scaling of the distance to the phase screen may be incorporated by a  $\sec^{5/6}(\theta)$ scaling of the scintillation variance for $b\lesssim r_{\rm F}$ \citep{wheelon2}. Combining these two effects, we conclude that the fractional scintillation variance scales with zenith angles of the source as $\sigma^2_{\rm fr}[\mv{b},\theta] \propto \sec^{11/6}(\theta)$, or approximately $\sigma_{\rm fr}[\mv{b},\theta] \propto \sec^{11/12}\theta$. The above contributions to zenith-angle scaling are summarised in Table \ref{ch4tab:za_scaling}. In Fig. \ref{ch4fig:za_dependence}, we have computed the fractional scintillation rms $\sigma_{\rm fr}[\mv{b},\theta]$ as a function of source zenith angle for various baseline lengths (left panel). We show curves for both the $\sec^{11/12}\theta$ scaling approximation, and for an accurate numerically computed ionospheric path-length and Fresnel-length increase at each zenith angle. For $b\lesssim 10r_{\rm F}$, the latter curves follow the expected $\sec^{11/12}\theta$ scaling for $\theta\lesssim 40$~degrees, but increase less rapidly than $\sec^{11/12}\theta$ for $\theta\gtrsim 40$~degrees.  Finally, in addition to the above effects, off-nadir viewing in the presence of curvature also results in non-zero angles of incidence on the ionosphere which in turn leads to refractive shift in the apparent position of sources. Though this is an important factor for self-calibration, since we are interesting in computing scintillation noise from an ensemble of sources (drawn from some source-counts), we will discount this refractive position shift.\\
%
%
%
%
%
%

%
%
%
%
\begin{table}
\centering
\caption{Summary of the approximate effects of off-zenith viewing on the fractional scintillation noise variance $\sigma^2_1[\mv{b},\theta]$  \label{ch4tab:za_scaling}}
\begin{tabular}{ll}
\hline
Factor & Approx. effect on $\sigma^2_{\rm fr}[\mv{b},\theta]$\\ \hline\\
Ionospheric path-length  & $\sec\theta$ \\
Distance to phase screen & $\sec^{5/6}\theta$ \\
Total & $\sec^{11/6}\theta$\\   \hline 
\end{tabular}
\end{table}
\subsection{Zenith-angle facets}
To compute the effects of scintillation noise for an arbitrarily large field of view, we will take a `facet' approach in conjunction with the above zenith-angle scaling. In this approach we are essentially dividing the sky into different `facets' and adding the scintillation noise from each facet in quadrature. Facets here refer to annuli at varying zenith angles. To do so, we have to first justify the implicit assumption that scintillation noise between sources in different facets is uncorrelated. Two factors affect the coherence of measured visibilities from different facets.
\begin{enumerate}
\item Angular decorrelation of scintillation noise as given in equation (\ref{ch4eqn:general_cov}). The angular coherence scale for scintillation is (as discussed before) $\Delta\mv{l}_{\rm sc}\approx 2r_{\rm F}/h$ for $b\lesssim r_{\rm F}$, and $\Delta\mv{l}_{\rm sc}\approx 2b/h$ for $b\gtrsim r_{\rm F}$. 
\item Geometric (or fringe) decorrelation of visibilities due to varying geometric delays between an ensemble of sources. If the facets are not sparsely populated by sources contributing to scintillation noise, then the angular separation over which we expect decorrelation is $\Delta\mv{l}_{\rm geo} \approx \lambda/b$. 
\end{enumerate}

It is straightforward to show that for $b\lesssim r_{\rm F}$, $\Delta\mv{l}_{\rm sc}<\Delta\mv{l}_{\rm geo}$, and the dominant source of decorrelation is the angular decorrelation of scintillation noise. For this short-baseline case, if the average separation between sources contributing to the sky power spectrum exceeds $\theta_{\rm F}=2r_{\rm F}/h$, then we are justified in using our faceted approach. At $150$~MHz, we have $\theta_{\rm F} \approx 7$~arcmin. We expect to find one source per $7\times7$~arcmin$^2$ of sky within $1$~dex of about $40$~mJy based on the source counts from equation (\ref{ch4eqn:dsc}). The majority of sources contributing to scintillation noise are well above this flux threshold, and their mutual separation safely exceeds $\theta_{\rm F}$. Hence they scintillate independently.\\

For $b\gtrsim r_{\rm F}$ the dominant source of decorrelation is geometric (or fringe) decorrelation: on the longer baselines the scintillation noise is coherent over an angular extent that is larger than the interferometer fringe spacing. The largest baselines on which one might expect to measure the $21$-cm power-spectrum with statistical significance in current and future instruments is about $b=1.5$~km. The fringe decorrelation scale for such a baseline at $150$~MHz of about $4.6$~arcmin. Again, we expect to find a source of flux within $1$~dex of about $20$~mJy at $150$~MHz in a $4.6\times4.6$~arcmin$^2$ area of the sky. This flux threshold is still significantly below the that of sources which contribute to the bulk of the observed scintillation noise. Hence even in the long baseline case, decomposing the sky into different facets, and summing up the scintillation noise from each facet in quadrature is justified.\\

\begin{figure*}
\centering
\includegraphics[width=\linewidth]{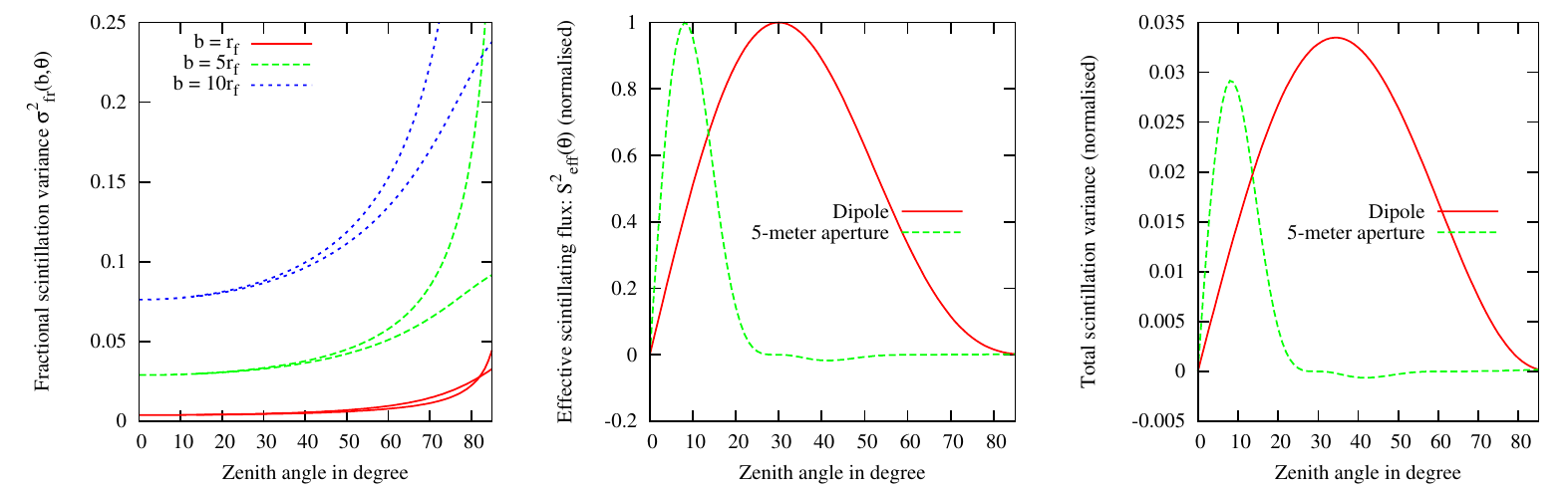}
\caption{Left panel: Scintillation noise variance at $150$~MHz on a $1$~Jy source as a function of zenith angle of the source, for baseline lengths of $1$, $5$ and $10$ times the Fresnel-length ($r_{\rm F}=310$~m). For each baseline length values, the curve with the steeper rise at larger zenith angles follows the $\sec^{11/6}\theta$ approximation, while the other curve results from an accurate numerical computation of zenith-angle scaling effects. Middle panel: The effective scintillating flux as a function of zenith angle for a dipole primary beam and a $5$~meter primary aperture. Right panel: Product of the curves from the other two panels for $b=r_{\rm f}$ showing the relative contribution to scintillation variance from different zenith angle segments. \label{ch4fig:za_dependence}}
\end{figure*}

Hence, to compute the scintillation noise variance from an arbitrarily large field of view, we do the following.
\begin{enumerate}
\item Decompose the sky in to annuli (or `facets') at varying zenith angle $\theta$. The solid angle within the annuli is given by $2\upi  \sin\theta {\rm d}\theta$
\item Compute the effective scintillating flux for sources within each annulus: ${\rm d}S^2_{\rm eff}(\theta)/{\rm d}\theta$ 
\item Sum the resulting scintillation noise variance values from each annulus while taking into account the zenith-angle scaling law shown in Fig. \ref{ch4fig:za_dependence} (left panel).
\end{enumerate}
The last step can be written as\footnote{We will absorb the factor $2\upi\sin\theta$ from the differential solid angle into the effective beam in equation (\ref{ch4eqn:beff_theta})} 
\begin{equation}
\label{ch4eqn:laststep}
\sigma^2[V(\mv{b})] = \int {\rm d}\theta\, \frac{{\rm d}S^2_{\rm eff}(\theta)}{{\rm d}\theta} \sigma_{\rm fr}^2[\mv{b},\theta] 
\end{equation} 
Using the source counts from equation (\ref{ch4eqn:dsc}), and following a procedure similar to \citet{speckle}, we can write the effective scintillating flux within a zenith-angle segment around $\theta$ as
\begin{equation}
\frac{{\rm d}S_{\rm eff}^2({\rm d}_{\rm prim},\nu,\theta)}{{\rm d}\theta} = \frac{C\nu^{-\beta}B_{\rm eff}(d_{\rm prim},\nu,\theta)S_{\rm max}^{3-\alpha}(d_{\rm prim},\nu)}{3-\alpha},
\end{equation}


%
where $S_{\rm max}(d_{\rm prim},\nu)$ is the flux density of the brightest source contributing to scintillation, $B_{\rm eff}$ is the effective beam area under a zenith angle segment at $\theta$, and is given by
\begin{equation}
\label{ch4eqn:beff_theta}
B_{\rm eff}(d_{\rm prim},\nu,\theta) = 2\upi \sin(\theta)\, B^{\alpha -1}(d_{\rm prim},\nu,\theta)\,\,\textrm{rad,}
\end{equation}
$B(d_{\rm prim},\nu,\theta)$ being the beam response of the primary antenna element. For an electrically short dipole element we will choose 
\begin{equation}
B_{\rm dip}(\nu,\theta) = \cos^2\theta,
\end{equation}
and for a circular aperture of diameter $d$, we use the usual Airy function
\begin{equation}
\label{ch4eqn:firststep}
B_{\rm circ}(\nu,\theta) =\left( \frac{2J_1(\upi d_{\rm prim} \sin(\theta)/\lambda)}{\upi d_{\rm prim} \sin(\theta)/\lambda}\right)^2.
\end{equation}
%

%
%

%
The middle-panel in Fig. \ref{ch4fig:za_dependence} shows the variation of ${\rm d}S^2_{\rm eff}(\theta)/{\rm d}\theta$ with $\theta$ for the case of a dipole and a circular aperture of $5$~meter diameter (the curves have been normalised to a maximum value of unity). We have used the typical values for the source counts from section \ref{ch4sec:basics}, which gives
\begin{eqnarray}
\frac{{\rm d}S_{\rm eff}^2(d_{\rm prim},\nu,\theta)}{d\theta} &\approx&  6\times 10^3\,\left( \frac{\nu}{150\textrm{ MHz}}\right)^{-0.8} \nonumber \\
&& \times \, \left(\frac{B_{\rm eff}(\nu,\theta)}{1\textrm{ rad}} \right)  \left( \frac{S_{\rm max}}{1\textrm{ Jy}}\right)^{0.5}\nonumber \\
&& \textrm{ Jy$^2$rad$^{-1}$} 
\end{eqnarray}
${\rm d}S^2_{\rm eff}(\theta)/{\rm d}\theta$ initially increases with $\theta$ due to an increase in the solid angle of annuli with zenith angle. For larger values of $\theta$ the curve falls off due to the rapidly decreasing primary beam gain away from zenith. For a dipole beam, most of the scintillating flux in the sky is around a zenith angle of $30$~degree. However when combined with the $\sec^{11/6}\theta$ scaling of the fractional scintillation variance most of the scintillation noise itself comes from zenith angles in the vicinity of $\theta\sim 45$~degree as seen in the right panel of Fig. \ref{ch4fig:za_dependence}. Fig. \ref{ch4fig:za_dependence} also shows that even for a modest $5$~meter wide aperture pointed towards zenith, most of the scintillation still comes from zenith angles $\theta\lesssim 20$~deg for which $\sigma_{\rm fr}^2[\mv{b},\theta] \propto \sec^{11/6}\theta$ is only about $10$\% higher than its value at zenith. \\

We thus conclude the following.
\begin{enumerate}
\item  Even for modest apertures ($\gtrsim 5$~meter diameter) pointed towards zenith, we can simply use the equations (\ref{ch4eqn:scint_fullps}) and (\ref{ch4eqn:scint_seff}) by neglecting zenith-angle scaling effects. This leads to an underestimate of scintillation noise rms of less than $1$\%.
\item For dipoles however, the above approximation will lead to a modest underestimation of scintillation noise rms of about $10$\%.
\item For apertures pointed off zenith, one must use the the numerically computed scaling law shown in Fig. \ref{ch4fig:za_dependence}. For zenith angles $\lesssim 40$~deg we can approximately scale the scintillation noise variance (computed for zenith) by $\sec^{11/6}\theta$.
\end{enumerate}
%
%
%
%
%
%
\section{Fourier synthesis effects}
\label{ch4sec:fse}
In this section, we describe the effects of Earth rotation and bandwidth synthesis on scintillation noise. For now, we assume that scintillation noise has not been mitigated by calibration. We discuss calibration effects in section \ref{ch4sec:cal}. We will also not use the exact forms for the time and frequency coherence functions which we derive later in Sections \ref{ch4sec:stcoh} and \ref{ch4sec:bline_migration} respectively, but rather make simplified calculations using the coherence time and bandwidth instead. We will use this `toy model' to develop understanding of the problem while at the same time providing fairly accurate results. Finally, we assume that wide-field effects have been accounted for in the gridding step via w-projection. \\

Earth rotation synthesis yields visibilities on a 3-dimensional grid in the $(u,v,\nu)$ domain. Since we are primarily concerned with $21$-cm power spectrum estimation, we will assume that the visibilities are gridded with uniform weights, that is, all visibilities falling within a $(u,v,\nu)$ cell are averaged\footnote{Imaging of point-like sources may use other `optimal' weighting schemes, but angular power spectrum estimation typically dictates the use of uniform weights.}. Within the temporal coherence time-scale, scintillation noise is correlated between disparate baselines that are averaged into the same $u,v$ cell (at different $\nu$). Hence estimating the spectral coherence of scintillation noise in the gridded visibilities essentially becomes a laborious book-keeping exercise. Fortunately, most current and future arrays fall into one of two limiting categories (discussed below) that allow us to make justified simplifications to alleviate the burden of book-keeping.   
\subsection{Minimally redundant arrays}
To ensure Nyquist sampling, the $uv$-plane grid resolution is usually chosen to be
\begin{equation}
\Delta u_{\rm cell} = \Delta v_{\rm cell} = \frac{d_{\rm prim}}{2\lambda},
\end{equation} 
where $d_{\rm prim}$ is the primary aperture diameter and $\lambda$ is the wavelength. A baseline of length $u=b/\lambda$ covers an arc of length $2\upi b/\lambda$ during $24$ hours of synthesis. Hence, it spends an amount of time equal to 
\begin{equation}
\label{ch4eqn:taucell}
\Delta\tau_{\rm cell}(b) = \frac{24\times3600}{2\upi}\frac{d_{\rm prim}}{2b} \,\,\, {\rm seconds}
\end{equation}
in a grid-cell. Similarly in the frequency domain, a baseline of length $b$ moves the length of a $uv$-cell within a frequency interval given by
\begin{equation}
\label{ch4eqn:nucell}
\Delta \nu_{\rm cell}(b) = \frac{d_{\rm prim}\nu}{2b}
\end{equation}

Hence, visibilities from a given baseline are integrated into a $uv$-cell over a time and frequency interval of $[\Delta\tau_{\rm cell},\Delta\nu_{\rm cell}]$. For a synthesis bandwidth of $\Delta\nu_{\rm syn}$, and a (baseline-length dependent) scintillation coherence timescale of $\Delta\tau_{\rm coh}$, we will call an array `minimally redundant' if the number of baselines contributing to a $uv$-cell of size $[\Delta u_{\rm cell},\Delta v_{\rm cell}]$ within an aperture synthesis interval of [$\Delta\tau_{\rm coh},\Delta\nu_{\rm syn}]$ is almost always unity or less. Under the above definition of redundancy, we find that on average only about $10$\% or less of LOFAR's core baselines and about $3-5$\% (frequency dependent) of MWA 128T baselines are deemed redundant: LOFAR and MWA fall under the category of minimally redundant arrays for scintillation noise purposes. However SKA-LOW, HERA, and PAPER are not in this regime (see section \ref{ch4sec:filled}).  \\

Due to the minimal redundancy assumption, we can simply disregard the coherence of scintillation noise between any pair of disparate baselines in our calculations since they will never be averaged together during aperture synthesis. Since scintillation noise is inherently broadband (for weak scintillation), the `monochromatic' thermal and scintillation noise contribution of a given baseline to a $uv$-cell can be written as.
\begin{equation}
\label{ch4eqn:scthnoise}
\sigma_{\rm th}(b) \approx \frac{{\rm SEFD}}{\sqrt{2 \Delta \nu_{\rm ch} \Delta\tau_{\rm cell}}} \textrm{ and } \sigma_{\rm sc}(b) \approx \frac{S_{\rm eff}\sigma_{\rm fr}[b]}{\sqrt{\Delta\tau_{\rm cell}/\Delta\tau_{\rm coh}(b)}}
\end{equation}
where $\Delta\nu_{\rm ch}$ is the integration bandwidth (or channel width) of the visibilities, and $\Delta\tau_{\rm cell}/\Delta\tau_{\rm coh}$ is the number of independent `scints' that are averaged into the $uv$-cell. It is important to note here that the minimum allowed value of $\Delta \tau_{\rm cell}/\Delta\tau_{\rm coh}(b)$ is unity, since one cannot have less than $1$ independent `scint' averaged into a cell. \\

\begin{figure*}
\includegraphics[width=\linewidth]{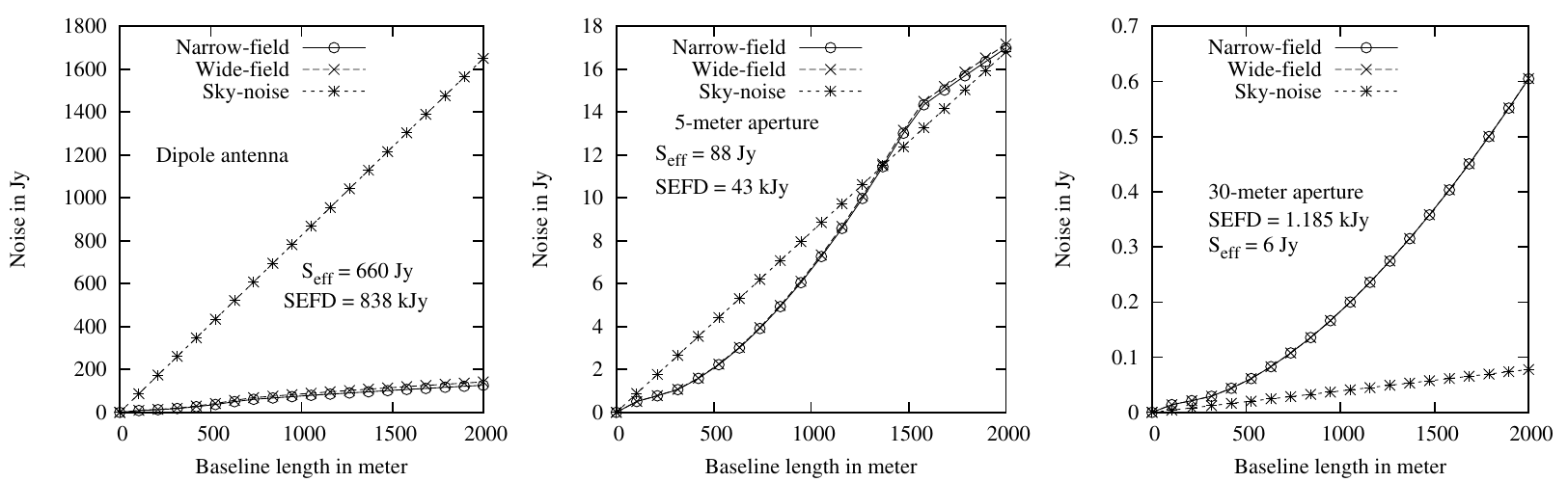}
\caption{Scintillation noise contribution of a single baseline to a $uv$-cell (monochromatic case) as a function of baseline length with and without taking widefield effects into account in scintillation noise calculations. The three panels are for a short-dipole primary antenna, and circular apertures of $5$ and $30$~meter diameter. The assumptions that went into computing the above figure are summarised in table \ref{ch4tab:nf_wf}. \label{ch4fig:nf_wf}}
\end{figure*}
%
To gauge the relative magnitudes of scintillation and thermal noise contribution from a single baseline, in Figure \ref{ch4fig:nf_wf}, we present the values of $\sigma_{\rm sc}(b)$ and $\sigma_{\rm th}(b)$. Though this is not done in practice, to compare the two noise values on equal footing, we have chosen a baseline dependent channel width of $\Delta\nu_{\rm ch}=\Delta\nu_{\rm cell}$ since this is the average frequency interval over which a baseline falls into a $uv$-cell. Hence we call this the `monochromatic' case, and we will account for frequency-coherence properly in section \ref{subsec:freq_coh}. Table \ref{ch4tab:nf_wf} summarises the assumptions that have gone into computations associated with Fig. \ref{ch4fig:nf_wf}.\\

\begin{table}
\centering
\caption{Assumptions for calculations leading to Fig. \ref{ch4fig:nf_wf}\label{ch4tab:nf_wf}}
\begin{tabular}{ll}
\hline
Quantity & Value\\ \hline
$r_{\rm diff}$ & 10~km\\
$\nu$ & 150~MHz\\
SEFD & equation (\ref{ch4eqn:sefdvals})\\
$S_{\rm eff}$ & equation (\ref{ch4eqn:seffvals})\\
$v$ & $500$~km~hr$^{-1}$\\
$\Delta\tau_{\rm cell}$ & equation (\ref{ch4eqn:taucell})\\
$\Delta\nu_{\rm ch}=\Delta\nu_{\rm cell}$ & equation (\ref{ch4eqn:nucell}) \\
$\Delta\tau_{\rm coh}$ & $2r_{\rm F}/v$ for $b<r_{\rm F}$\\
& and $2b/v$ for $b>r_{\rm F}$\\
$\sigma_{\rm fr}[b]$ & equation (\ref{ch4eqn:scint_seff})\\ \hline
\end{tabular}
\end{table}

In Fig. \ref{ch4fig:nf_wf}, the curves marked `wide-field' and `narrow-field' were computed with and without accounting for the zenith-angle scaling of scintillation respectively. As expected, notwithstanding a small difference for the dipole case, the two curves are practically indistinguishable. More importantly, Fig. \ref{ch4fig:nf_wf} shows that scintillation noise is larger for smaller receiving apertures. For an aperture of $d_{\rm prim}\gtrsim5$~meter, the monochromatic scintillation noise contribution from a single baseline is of the same magnitude of larger than thermal noise. The dominance of scintillation noise over thermal noise is more pronounced for larger primary apertures since the sky noise scales with the aperture diameter as $d^{-2}$ whereas the effective scintillating flux scales as $d^{-1.5}$ (see equations (\ref{ch4eqn:sefd}) and (\ref{ch4eqn:seff})). Finally, the `break' in the scintillation noise curve for $d_{\rm prim}=5$~m (middle panel) around $b=1500$~m is due to our assertion of a minimum bound of unity for $\Delta\tau_{\rm cell}/\Delta\tau_{\rm coh}$. In this case for $b\gtrsim 1500$~m, the baselines spend less time in a $uv$-cell during synthesis than the typical scintillation coherence timescale (for $v=500$~km~hr$^{-1}$).  
\subsection{Frequency coherence and the delay transform}
\label{subsec:freq_coh}
The inherent spectral coherence of scintillation in the weak scattering regime is given by $\Delta\nu/\nu\sim1$. Hence in practice, spectral decorrelation of scintillation in the $uv$-plane is dominated by the natural migration of baselines owing to stretching of baseline length in wavelength units with frequency. Note that Earth rotation synthesis is typically employed in EoR experiments to `fill-up' the $uv$-plane. This does not ensure spectral coherence of \emph{measured} scintillation because scintillation is not expected to be coherent over timescales on which Earth rotation synthesis is performed (several hours). Hence the same $uv$-cell if sampled (by different baselines) at two different frequency channels at different times during the synthesis will invariably have incoherent scintillation noise realisations at the two frequency channels. The relevant $uv$-coverage to consider for scintillation noise calculations is the snapshot $uv$-coverage at different frequencies.
\subsection{Baseline migration in the delay-domain}
\label{ch4sec:bline_migration}
The snapshot $uv$-coverage is array dependent, but as argued earlier, we expect a given baseline of length $b$ to cross a $uv$-cell in a frequency interval of $\Delta\nu_{\rm cell} = d_{\rm prim}\nu/(2b)$. For a minimally redundant array, $\Delta\nu_{\rm cell}$ sets the frequency scale over which the measured scintillation noise decorrelates. More formally, if we assume that visibilities are measured with an integration bandwidth of $\Delta\nu_{\rm ch}$, and that there are $N_{\rm ch}$ such continuous channels forming a synthesis bandwidth of $\Delta\nu_{\rm syn}=N_{\rm ch}\Delta\nu_{\rm ch}$, then for each $uv$-cell, we can define a normalised frequency coherence function for scintillation noise as
\begin{equation}
\label{ch4eqn:rij}
R_{\rm sc}[\nu_i,\nu_j] = \frac{N_{ij}}{\sqrt{N_iN_j}}
\end{equation}
where $N_i$ and $N_j$ are the number of visibilities that fall into the $uv$-cell at channel $i$ and $j$ respectively, and $N_{ij}$ are the number of visibilities from the same baseline that fall into the $uv$-cell at both the channels. Due to the natural migration of baselines with frequency, for a minimally redundant array, we expect $R_{\rm sc}[\nu_i,\nu_j]$ to reach a value of $0.5$ for a frequency separation of $\pm\Delta\nu_{\rm cell}/2$:
\begin{equation}
\label{ch4eqn:rijwidth}
|\nu_i-\nu_j| = \frac{\Delta\nu_{\rm cell}}{2} = \frac{d_{\rm prim}(\nu_1+\nu_2)/2}{4b}\textrm{ when }R_{\rm sc}[\nu_i,\nu_j]\approx 0.5
\end{equation}

\begin{figure}
\centering
\includegraphics[width=\linewidth]{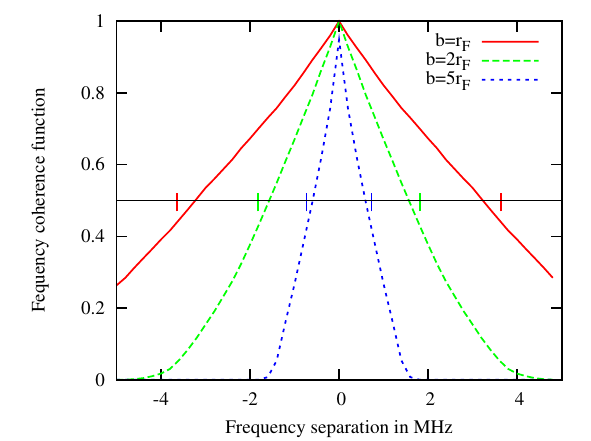}
\caption{Frequency coherence function of observed visibility scintillation (normalised to unity) for different baseline lengths. The different curves have been numerically evaluated by following the migration of LOFAR baselines with frequency during a synthesis of $12$~hours on the NCP. The small bars on the $y=0.5$ line show the points for each baseline given by $[-d_{\rm prim}\nu/(4b)\,\,\, d_{\rm prim}\nu/(4b)]$ which is the (approximate) expected frequency coherence width.\label{ch4fig:freq_coherence}}
\end{figure}

In Fig. \ref{ch4fig:freq_coherence}, we plot the numerically computed frequency coherence function $R_{\rm sc}[i,j]$ for the case of NCP observations with LOFAR. The curves in the figure were computed using equation (\ref{ch4eqn:rij}) where $N_i$, $N_j$ and $N_{ij}$ were evaluated by tracing all LOFAR baselines as they were gridded into the $uv$-cells during $12$~hours of synthesis on the North Celestial Pole (NCP) field. The figure also shows the expected decorrelation bandwidth computed from equation (\ref{ch4eqn:rijwidth}) (vertical bars), which gives a fairly accurate expression for the coherence bandwidth of measured visibility scintillation. More importantly, the curves in Fig. \ref{ch4fig:freq_coherence} have an approximately linear linear drop as a function of frequency separation. This is due to the fact that in our numerical calculations, we employed nearest neighbour gridding which is similar to gridding by convolution with a top-hat convolution kernel. The curves in Fig. \ref{ch4fig:freq_coherence} are thus the autocorrelation function of a top-hat function in 2 dimensions ($u$ and $v$) which is expected to have a conical shape given by \citep{fried1967}
\begin{eqnarray}
\label{ch4eqn:triangle}
R_{\rm sc}[\nu_i,\nu_j]  &=& \frac{2}{\upi}\left[ \cos^{-1}\left(\frac{\Delta\nu}{\Delta\nu_{\rm cell}}\right) \right. \nonumber \\
&& \left. -\,\frac{\Delta\nu}{\Delta\nu_{\rm cell}}\sqrt{1-\left(\frac{\Delta\nu}{\Delta\nu{\rm cell}}\right)^2} \right];\, \Delta\nu<\Delta\nu_{\rm cell}\nonumber \\
&=& 0\,\,\,{\rm otherwise}
\end{eqnarray}
where $\Delta\nu=|\nu_i-\nu_j|$. We note here that $R_{\rm sc}$ should be evaluated as the autocorrelation function of the particular kernel-function being used in the gridding by convolution.\\

Since the observed frequency of the $21$-cm signal corresponds to line of sight distance (over small bandwidths), $21$-cm power spectrum estimation involves a Fourier transform of gridded visibilities along the frequency axis at each $uv$-cell. This transforms the frequency axis into a delay ($\eta$) axis, and essentially casts the gridded data on the cosmological wavenumber space on all three dimensions\footnote{Hence the name: delay transform. Delay $\eta$ approximately corresponds to line of sight wavenumber}. The functional behaviour of scintillation noise in the $\eta$ domain is then given by the Fourier transform of $R_{\rm sc}[\nu_i,\nu_j]$, which may be approximated by the sinc(.)$^2$ function. However, it is possible to have cases wherein $\Delta\nu_{\rm cell}>\Delta\nu_{\rm syn}$ such that the frequency coherence function $R_{\rm sc}$ is essentially multiplied by a top-hat window of width $\Delta\nu_{\rm syn}$ (see the $b=r_{\rm F}$ curve in Fig. \ref{ch4fig:freq_coherence} for instance). This yields an additional convolution in the delay domain with a sinc function. The Fourier transform of the coherence function $R_{\rm sc}$ can hence be approximately\footnote{We have approximated $R_{\rm sc}[\nu_i,\nu_j]$ of equation (\ref{ch4eqn:triangle}) with a triangular function, which despite being convenient, gives slightly elevated sidelobe levels in $\eta$ space (Gibbs' ringing)} written as
\begin{equation}
\label{ch4eqn:reta}
\widetilde{R}(\eta) = \frac{\Delta\nu_{\rm cell}}{\Delta\nu_{\rm syn}} \left(\frac{\sin\left(\upi\eta\Delta\nu_{\rm cell}\right)}{\upi\eta\Delta\nu_{\rm cell}}\right)^2 \ast \left(\Delta\nu_{\rm syn}\frac{\sin\left(\upi\eta\Delta\nu_{\rm syn}\right)}{\upi\eta\Delta\nu_{\rm syn}}\right)
\end{equation}
where $\ast$ is the convolution operator, and the factor $\Delta\nu_{\rm syn}$ has been absorbed into the convolution kernel for easy interpretation: for $\Delta\nu_{\rm syn}\gg\Delta\nu_{\rm cell}$ the convolution kernel varies rapidly in $\eta$, and the area under the kernel is unity. Hence for $\Delta\nu_{\rm cell}\lesssim\Delta\nu_{\rm syn}$, the multiplication by the top-hat window is inconsequential, and the above equation reduces to
\begin{equation}
\label{ch4eqn:sincfilter}
\widetilde{R}(\eta) = \frac{\Delta\nu_{\rm cell}}{\Delta\nu_{\rm syn}}\left(\frac{\sin(\upi\eta\Delta\nu_{\rm cell})}{\upi\eta\Delta\nu_{\rm cell}}\right)^2\,\,\,\,\Delta\nu_{\rm cell}\lesssim \Delta\nu_{\rm syn.}
\end{equation}
Hence the scintillation noise contribution of a single baseline in the $u,v,\eta$ domain is
\begin{equation}
\sigma^2_{\rm sc}[b,\eta] = \widetilde{R}(\eta)\sigma^2_{\rm sc}[b],
\end{equation}
which on using equations (\ref{ch4eqn:scthnoise}) and (\ref{ch4eqn:reta}) becomes
\begin{eqnarray}
\label{ch4eqn:fullsigmasc}
\sigma^2_{\rm sc}[b,\eta] &=& \frac{S^2_{\rm eff}\sigma^2_{\rm fr}[b]\Delta\nu_{\rm cell}/\Delta\nu_{\rm syn}}{\Delta\tau_{\rm cell}/\Delta\tau_{\rm coh}(b)}\left(\frac{\sin(\upi\eta\Delta\nu_{\rm cell})}{\upi\eta\Delta\nu_{\rm cell}}\right)^2 \nonumber \\
&&\ast\, \left(\upi\Delta\nu_{\rm syn}\frac{\sin\left(\upi\eta\Delta\nu_{\rm syn}\right)}{\upi\eta\Delta\nu_{\rm syn}}\right) \nonumber \\
&\approx&\frac{S^2_{\rm eff}\sigma^2_{\rm fr}[b]\Delta\nu_{\rm cell}/\Delta\nu_{\rm syn}}{\Delta\tau_{\rm cell}/\Delta\tau_{\rm coh}(b)}\left(\frac{\sin(\upi\eta\Delta\nu_{\rm cell})}{\upi\eta\Delta\nu_{\rm cell}}\right)^2,\nonumber \\
&&\,\,\textrm{for } \Delta\nu_{\rm cell}\lesssim \Delta\nu_{\rm syn,}
\end{eqnarray}
where we will evaluate $\Delta\nu_{\rm cell}$ at the centre frequency (say $\nu_0$) within the synthesis bandwidth. We again caution the reader that the maximum permissible value of $\Delta\tau_{\rm cell}/\Delta\tau_{\rm coh}$ is unity, since one cannot have less than one independent `scint' within an integration epoch. In addition, the convolution in equation (\ref{ch4eqn:fullsigmasc}) is difficult to compute numerically due to a large support in $\eta$ for the sinc functions. Hence, an easier and more accurate method to compute $\sigma^2_{\rm sc}$ is to numerically evaluate the Fourier transform of $R_{\rm sc}$ from equation (\ref{ch4eqn:triangle}) with the relevant truncation for the case of $\Delta\nu_{\rm cell}>\Delta\nu_{\rm syn}$. \\
\subsection{Thermal noise in the delay-domain}
Evaluation of the thermal noise contribution is relatively straightforward. Since thermal noise is uncorrelated between frequency channels, we can define the thermal noise frequency coherence function as
\begin{equation}
R_{\rm th}[\nu_i,\nu_j] = \delta_{ij}
\end{equation}
where $\delta_{ij}$ is the Kronecker-delta function. Taking the delay transform of $R_{\rm th}[\nu_i,\nu_j]$ we can write the thermal noise contribution of a given baseline to a $uv$-cell in the delay domain as
\begin{equation}
\sigma^2_{\rm th}[\eta,b] = \frac{1}{N_{\rm ch}}\sigma^2_{\rm th}[b],
\end{equation}
which on using equation (\ref{ch4eqn:scthnoise}) becomes
\begin{equation}
\label{ch4eqn:fullsigmath}
\sigma^2_{\rm th}[\eta,b] =\frac{{\rm SEFD}^2}{2\Delta\nu_{\rm syn}\Delta\tau_{\rm cell}}
\end{equation}

Equations (\ref{ch4eqn:fullsigmasc}) and (\ref{ch4eqn:fullsigmath}) give the scintillation noise and thermal noise contribution to a $u,v,\eta$ cell from a single baseline. Due to the minimally redundant assumption both $\sigma^2_{\rm sc}$ and $\sigma^2_{\rm th}$ will both be reduced in a full synthesis by the number of baselines that pass though the given $uv$-cell, and as such, their ratio is expected to still be given by $\sigma^2_{\rm sc}[\eta,b]/\sigma^2_{\rm th}[\eta,b]$. Since considerable effort has already been spent by various authors in computing the thermal noise contribution to the $21$-cm power spectra measured by various minimally redundant telescopes, we will present our results as the scintillation to thermal noise ratio.
\subsection{Scintillation to thermal noise ratio}
\begin{figure*}
\centering
\includegraphics[width=0.85\linewidth]{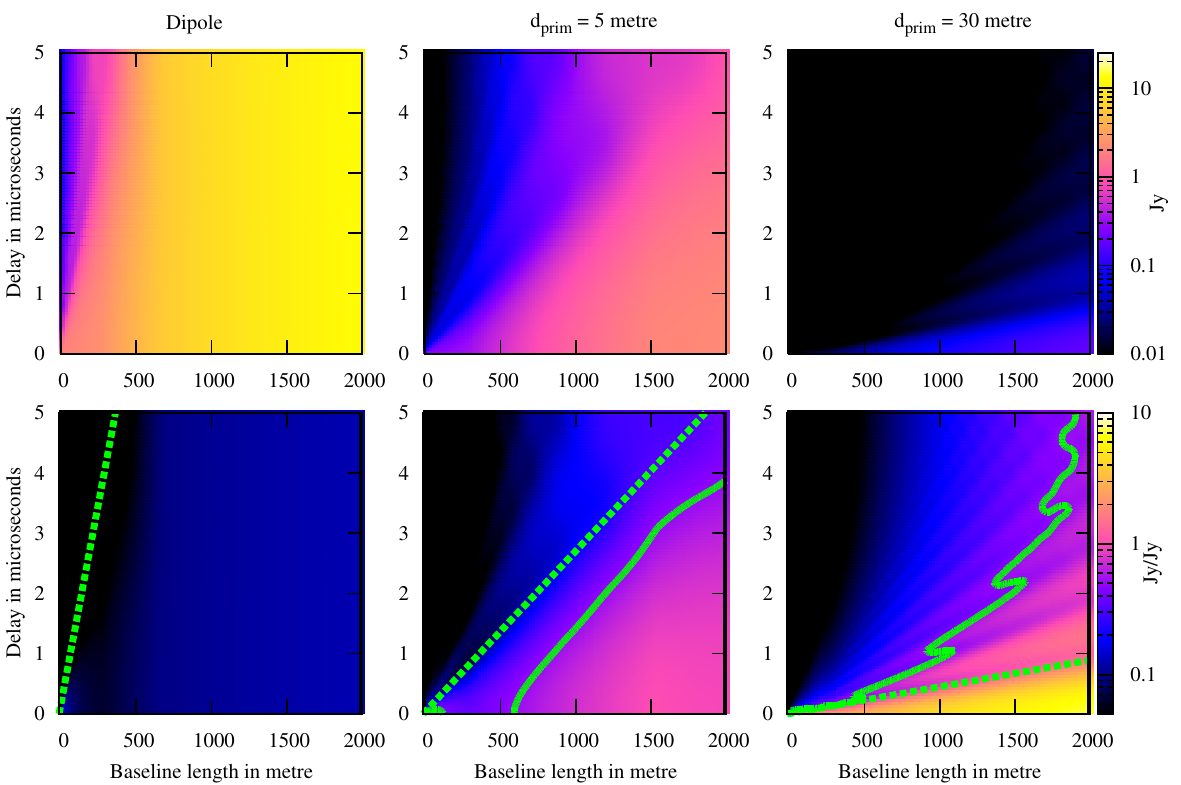}
\caption{Top row: scintillation noise in delay-baseline space evaluated using equation (\ref{ch4eqn:fullsigmasc}) for parameters summarised in table \ref{ch4tab:nf_wf} for the three cases of a dipole, $5$~m and $30$~m primary apertures. Bottom row: the corresponding ratio of scintillation and thermal noise (evaluated using equation (\ref{ch4eqn:fullsigmath})). The iso-contour line marks a ratio of $1/2$ that corresponds to $25$\% more integration than previously thought (thermal noise alone) to achieve the same power spectrum uncertainty. The broken line traces the sidelobe wedge \citep{vedantham2012}, above which scintillation noise can be partly mitigated using a suitable window function.\label{ch4fig:nf_wf_eta}}
\end{figure*}

Fig. \ref{ch4fig:nf_wf_eta} shows the both the scintillation noise and the ratio of scintillation to thermal noise for the same three cases of dipole, $d_{\rm prim}=5$~m, and $d_{\rm prim}=30$~m as in Fig. \ref{ch4fig:nf_wf}, whereas now, we have taken the frequency coherence of noise into account. In computing the values in Fig. \ref{ch4fig:nf_wf_eta}, we have made the same assumptions as in table \ref{ch4tab:nf_wf} along with $\Delta\nu_{\rm syn}=10$~MHz\footnote{Note that our choice of $\Delta\nu_{\rm syn}$ merely sets the resolution along the delay ($\eta$) space and does not affect the values plotted in Fig. \ref{ch4fig:nf_wf_eta}}. From Fig. \ref{ch4fig:nf_wf_eta}, we conclude that under the assumption summarised in table \ref{ch4tab:nf_wf}, (i) for a dipole primary aperture, scintillation noise is typically smaller than sky noise, (ii) for $d_{\rm prim}=5$~m, scintillation noise considerably lowers the power spectrum uncertainty in a region to the to right and bottom of the iso-contour line $\eta_{\mu{\rm sec}} = 3(b_{\rm km}-1/2)$ placed at a ratio of $1/2$. We have chosen this iso-contour since a ratio of $1/2$ requires $25$\% larger integration time to reach the same noise levels as computed previously in the absence of scintillation noise. Note also that sidelobes of the sinc function (see equation (\ref{ch4eqn:sincfilter})) are prominent in Fig. \ref{ch4fig:nf_wf_eta}. These sidelobes (not the main lobe) can be partly mitigated in the region defined by $\eta>2b/(d_{\rm prim}\nu)$ via a judiciously chosen window function prior to applying the delay transform. The line $\eta=2b/(d_{\rm prim}\nu)$ is shown in the figure as a broken line and marks the boundary of the well known sidelobe `wedge' in delay space\footnote{The boundary or `sidelobe horizon' here corresponds to a sources at the first null of the primary beam.} \citep{vedantham2012}. While the equations in this section and Fig. \ref{ch4fig:nf_wf} demonstrate how the scintillation noise power spectrum (in delay-baseline space) can be calculated for a generic instrument, in section \ref{ch4sec:snps} we will provide scintillation to thermal noise ratio estimates for different redshifts (or frequencies) and  ionospheric conditions, for some instrument specific parameters.
\subsection{Maximally redundant compact arrays}
\label{ch4sec:filled}
We have thus far considered sparse arrays with very few number of redundant baselines. Under these assumptions the scintillation noise in gridded $uv$-data after Fourier and Earth rotation synthesis can be computed by our knowledge of the temporal and spectral coherence of scintillation noise alone. This is however not the case for dense arrays with high filling factors such as the current LWA and PAPER, and the proposed NenuFAR, HERA and SKA (central core only) which have many redundant baselines which are averaged together and will invariably have correlated scintillation noise. Calculation of scintillation noise in this case becomes cumbersome since one needs to keep track of all the mutual coherence values for redundant baselines that are averaged into the same $uv$-cell at any given time/frequency interval. We can however circumvent this `book-keeping' for the case of fully filled arrays which are almost wholly within a Fresnel-scale. This is the case for proposed arrays such as HERA and SKA-LOW which have nearly fully filled apertures with a diameter of about $300$~meter ($r_{\rm F}=310$~meter at $150$~MHz). Since scintillation noise is coherent on all redundant baselines whose mutual separation does not exceed $r_{\rm F}$, we can proceed with scintillation noise calculations as follows.\\

Consider a fully filled (or maximally redundant)  array of diameter $d_{\rm core}\leq 2r_{\rm F}$. Let the diameter of each primary antenna element in the array be $d_{\rm prim}$. The number of interferometer elements in such an array will be $N_{\rm prim} = (d_{\rm core}/d_{\rm prim})^2$. The autocorrelation function (normalised to maximum value of unity) of a circular aperture of diameter $d_{\rm core}$ is given by \citep{fried1967}
\begin{equation}
R_{\rm core}(b) = \frac{2}{\upi}\left[ \cos^{-1}\left(\frac{b}{d_{\rm core}}\right)-\frac{b}{d_{\rm core}}\sqrt{1-\left(\frac{b}{d_{\rm core}}\right)^2} \right].
\end{equation}
With this normalisation, we can show that the area under $R_{\rm core}(d)$ is equal to $A_{\rm core}=\upi d^2_{\rm core}/4$. 
Since the total number of baselines is $N^2_{\rm prim}$, the baseline density function must be 
\begin{equation}
\Sigma_{\rm core}(b)=R_{\rm core}(b)N^2_{\rm prim}/A_{\rm core}.
\end{equation}
 With a primary aperture diameter of $d_{\rm prim}$, we must choose a $uv$-cell of dimensions $d_{\rm prim}/2 \times d_{\rm prim}/2$ for Nyquist sampling of the visibilities. Hence the number of baselines that will be coherently integrated (in a snapshot) within a $uv$-cell as a function of baseline length $b$ is
\begin{equation}
N_{\rm base}(b) = \frac{d^2_{\rm prim}}{4}\Sigma_{\rm core}(b) = \frac{d^2_{\rm core}}{\upi d^2_{\rm prim}} R_{\rm core}(b)
\end{equation}
Using this the thermal noise per $uv$-cell can be written as
\begin{equation}
\sigma_{\rm th}(b) = \frac{{\rm SEFD}}{\sqrt{2\Delta\nu\Delta\tau_{\rm coh}N_{\rm base}(b)}}
\end{equation}
On expanding the SEFD in terms of the sky temperature $T_{\rm sky}(\nu)$, we get
\begin{equation}
\sigma_{\rm th}(b) = \frac{ 8kT_{\rm sky} }{ d_{\rm prim}d_{\rm core} \sqrt{2\upi\Delta\nu\Delta\tau_{\rm coh}R_{\rm core}(b)} }
\end{equation}
where $k$ is the Boltzmann's constant. Note that we have chosen the integration time-scale to be the scintillation coherence time-scale to compare thermal and speckle noise on equal footing. However unlike the sparse array case where the associated bandwidth for thermal and scintillation noise calculations are set by baseline migration, for a filled aperture this is not the case since there are no `holes' in the $uv$-plane even for a snapshot in time. Hence the choice of $\Delta\nu$ is somewhat arbitrary, and we choose it to be $1$~MHz since this is the bandwidth within which we expect the $21$-cm signal to remain mostly coherent. Given the coherence of scintillation on all redundant baselines, the scintillation noise per $uv$-cell within a decorrelation time-scale is \citep{speckle}
\begin{eqnarray}
\sigma_{\rm sc}[b] &=& S_{\rm eff}\sigma_{\rm fr}[b]\\
S_{\rm eff} &\approx&  5.86 \left( \frac{d_{\rm prim}}{30\textrm{ m}}\right)^{-1.5} \left( \frac{\nu}{150\textrm{ MHz}}\right)^{-2.025}\textrm{ Jy}
\end{eqnarray}
Hence thermal noise scales as $d^{-2}_{\rm prim}$, but the effective scintillating flux (and hence the scintillation noise) scales as $S_{\rm eff}\propto d^{-1.5}_{\rm prim}$. The total scintillation noise per $uv$-cell decreases less rapidly than thermal noise with increasing primary antenna element size.\\

\begin{figure}
\centering
\includegraphics[width=\linewidth]{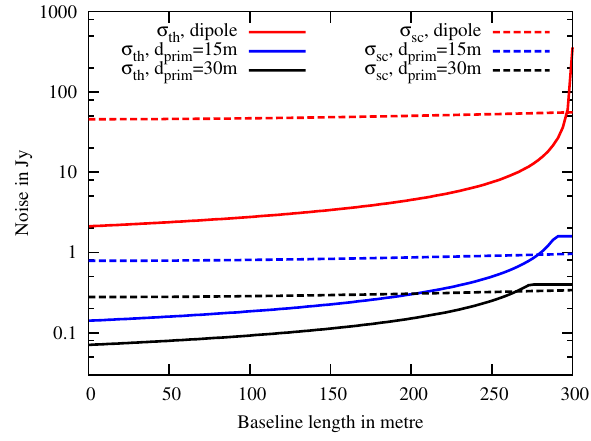}
\caption{Expected thermal noise (sky-noise only) plotted as solid lines, and scintillation noise plotted as dashed lines, as a function of baseline length for the case of a fully filled aperture of diameter $d_{\rm core}=300$~meter. The three pairs of curves are for cases wherein the total collecting area of $A_{\rm core}=\upi d^2_{\rm core}/4$ is spanned by primary antenna elements made of dipoles, and circular apertures of diameter $d_{\rm prim}=15$~meter and $d_{\rm prim}=30$~meter. Thermal noise curves have been computed for an integration bandwidth of $1$~MHz and integration time equal to the scintillation coherence time-scale $\Delta\tau_{\rm coh}\approx 4.46$~s. Thermal noise curves saturate at longer baselines when the number of baselines falling into a $uv$-cell approaches unity.\label{ch4fig:filled}}
\end{figure}

Figure \ref{ch4fig:filled} shows the computed values for scintillation and thermal noise for three different primary apertures: dipoles, and circular apertures of diameters of $d_{\rm prim}=15,\,30$~meter.
 In all cases we have assumed a filled array of diameter $d_{\rm core}=300$~m, so as to approximately reflect the cases of (i) LWA, SKA, and NenuFAR (dipole apertures), (ii) the proposed HERA telescope ($d_{\rm prim}=15$~m), and (iii) LOFAR augmented with extra high-band stations that fill the entire superterp\footnote{`Superterp' refers to the central dense part of the LOFAR array that currently has $12$ primary apertures with $d_{\rm prim}\sim 30$~m.}. Both telescopes will be dominated by scintillation noise. However, a critical advantage of filled arrays fully within the Fresnel scales comes from the fact that since there are no `holes' in the snapshot $uv$-coverage, and since scintillation noise is coherent over all baselines, there is no appreciable spectral decorrelation of scintillation noise due to migration of baselines on the $uv$-plane as a function of frequency. Hence the measured frequency coherence of scintillation noise is equal to the inherent spectral coherence-width for weak scintillation which is $\Delta\nu/\nu\sim 1$. Since scintillation noise is correlated across the compact core, scintillation for the fully filled aperture case is a source dependent (baseline independent) broadband random modulation of flux density. Due to this, we expect scintillation noise in filled arrays to be largely mitigated along with spectrally smooth Galactic and Extragalactic emission in the foreground subtraction step\footnote{The same is true for the current PAPER array wherein scintillation noise is filtered along with foregrounds (and part of the $21$-cm signal) in the delay domain.}. 
%
%
%
%
%
\section{Calibration effects}
\label{ch4sec:cal}
Self-calibration is typically employed in radio-interferometric data processing to remove among other corruptions, ionospheric effects. To understand residual ionospheric corruptions post-calibration, one has to evaluate the extent to which such effects are mitigated in the time, direction, and baseline dimensions. For instance, a self-calibration solution cadence of $t_{\rm sol}$ will be ineffective in mitigating visibility scintillation on timescale much smaller than $t_{\rm sol}$. Similarly, solutions obtained on a source in direction $\mv{l}_{\rm sol}$ may not fully mitigate corruptions on a source at position $\mv{l}$ if $|\mv{l}-\mv{l}_{\rm sol}|$ is larger than the angular coherence scale for visibility scintillation. Finally, since visibility scintillation effects are baseline dependent, calibration obtained from a set of baselines (only long-baselines for instance) may not be effective in mitigating the effects on a disparate set of baselines. Since different experiments may use varying data-processing strategies, we will proceed by discussing the impact of each of the above factors separately, and then proceed to compute scintillation noise by making representative assumptions about such strategies.
\subsection{The temporal coherence function}
\label{ch4sec:stcoh}
\begin{figure}
\centering
\includegraphics[width=\linewidth]{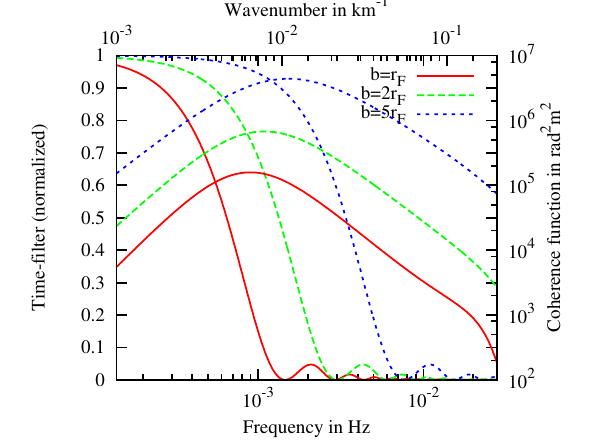}
\caption{Plot showing the time coherence function for visibility scintillation on different baselines (top and left axes), and the filtering function due to averaging of visibilities within a fringe decorrelation time-scale of $\Delta\tau_{\rm cell}$ from equation (\ref{ch4eqn:taucell}) (left and bottom axes). The fractional scintillation noise variance for each baseline is the area under the product of the coherence function and filtering function. \label{ch4fig:time_coherence}}
\end{figure}

During Fourier and Earth rotation synthesis, visibilities from a baseline are averaged\footnote{Note that post synthesis, the averaging interval of visibilities from the correlator ($\ll \Delta\tau_{\rm cell}$) is inconsequential to noise calculations.} over an interval of $\Delta\tau_{\rm cel}$. Both time averaging of visibilities and self-calibration mitigate scintillation noise on different time-scales, and hence the effect of both processes must be treated simultaneously. Temporal averaging over an interval of $\Delta\tau_{\rm cell}$ seconds, can be analytically expressed as a convolution of the observed visibilities with a square window function:
\begin{eqnarray}
V_{\rm avg}(\mv{b},t) &=& \int {\rm d}\tau V_{\rm M}(\mv{b},\tau) h_{\rm avg}(t-\tau)\nonumber \\
&=& V_{\rm M}(\mv{b},t)\ast h_{\rm avg}(t)
\end{eqnarray}
where $\ast$ is the convolution operator and
\begin{equation}
h_{\rm avg}(t) = \begin{cases}
\Delta\tau_{\rm cell}^{-1} & \textrm{if } -\Delta\tau_{\rm cell}/2 < t < \Delta\tau_{\rm cell}/2 \\
0 & \textrm{otherwise}
\end{cases}
\end{equation}
Note that though the true visibility varies with time, within an averaging interval of $\Delta\tau_{\rm cell}$ one can safely assume that this variation is small. Using the Fourier-convolution theorem we can write
\begin{equation}
\widetilde{V}_{\rm avg}(\mv{b},f) = \widetilde{V}_{\rm M}(\mv{b},f) \widetilde{h}_{\rm avg}(f),
\end{equation}
where frequency $f$ and time $t$ are Fourier conjugates, and $\widetilde{h}_{\rm avg}(f)$ is given by the sinc function
\begin{equation}
\widetilde{h}_{\rm avg}(f) = \frac{\sin(\upi \Delta\tau_{\rm cell}f)}{\upi \Delta\tau_{\rm cell} f}.
\end{equation}
The variance of the averaged visibilities can be computed in the Fourier domain as
\begin{equation}
\sigma^2[V_{\rm avg}(\mv{b})] = \int {\rm d}f |\widetilde{V}_{\rm M}(\mv{b},f)|^2 |\widetilde{h}_{\rm avg}(f)|^2
\end{equation}
Using Parseval's theorem, we can compute the variance of $V_{\rm avg}(\mv{b},t)$ in the Fourier domain as
\begin{equation}
\label{ch4eqn:pars}
\sigma^2[V_{\rm avg}(\mv{b},t)] = \int {\rm d}f \sigma^2[\widetilde{V}_{\rm M}(\mv{b},f)]  \left|\widetilde{h}_{\rm avg}(f) \right|^2
\end{equation}
where $\sigma^2[\widetilde{V}_{\rm M}(\mv{b},f)]$ is the Fourier transform of the temporal coherence function of scintillation from equation (\ref{ch4eqn:general_cov}). Assuming the sky power spectrum to be $S^2_{\rm eff}$, we can write
\begin{eqnarray}
\label{ch4eqn:general_tcoh}
\sigma^2[V_{\rm avg}(\mv{b})] &=& 4S^2_{\rm eff}\int {\rm d}^2\mv{q}\ips{\mv{q}} \sin^2(\upi\lambda h \mv{q}^2-\upi\mv{q}\bcdot\mv{b})\nonumber \\
&& \times \, \left[ \frac{\sin(\upi \Delta\tau_{\rm cell}\mv{q}\bcdot\mv{v})}{\upi \Delta\tau_{\rm cell} \mv{q}\bcdot\mv{v}}\right]^2,
\end{eqnarray}
since the Fourier transform of equation (\ref{ch4eqn:general_cov}) with respect to time yields a factor $\delta(f-\mv{q}\cdot\mv{v})$ which when integrated over frequency as in equation (\ref{ch4eqn:pars}) extracts the integrand at $f=\mv{q}\cdot\mv{v}$. The value of the integral in equation (\ref{ch4eqn:general_tcoh}) in general depends on the angle between $\mv{b}$ and $\mv{v}$, but for illustration, we consider a 1-dimensional scenario ($q_y=0$) where the two vectors are parallel. In Fig. \ref{ch4fig:time_coherence} we plot the two important factors in the integral namely (i) the product of the ionospheric power spectrum and the Fresnel-baseline filter ($\sin^2$ term) and (ii) the Fourier transform of the square window function (sinc$^2$ term). As seen in the figure, small scale turbulence gives scintillation with shorter time coherence (larger frequency) and is mitigated by averaging (sinc$^2$ function has small value). Hence averaging effectively removes contribution from small scale turbulence, but is ineffective in removing the contribution of large scale turbulence that is coherent over time-scales larger than the averaging interval.
\subsection{Solution cadence}
\label{ch4sec:solcad}
\begin{figure}
\centering
\includegraphics[width=\linewidth]{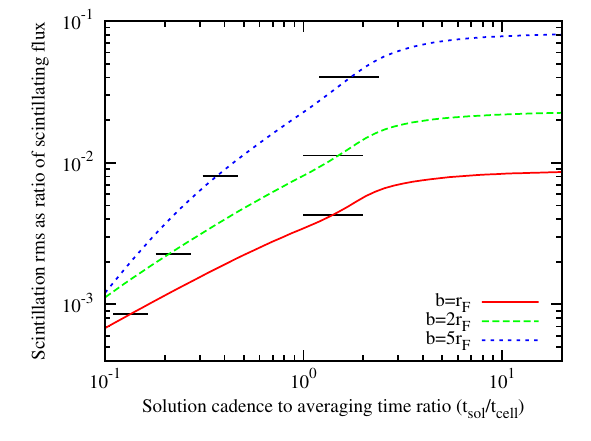}
\caption{Residual scintillation noise as a function of ratio between the solution cadence and averaging interval for different baseline lengths. The black horizontal lines has been placed at half and one-tenth of the peak values (attained at $t_{\rm sol}/t_{\rm avg}\gg 1$) to guide the eye.\label{ch4fig:opt_tsol}}
\end{figure}

Although not always valid, we first assume that sufficient number of constraints exist to mitigate scintillation noise from an arbitrary number of directions. If such self-calibration solutions are obtained every $t_{\rm sol}$ seconds, and all scintillating sources are subtracted using solutions in their respective directions, then taking visibility averaging also into account, we can write the residual visibility as
\begin{equation}
\Delta V_{\rm C}(\mv{b},t) = V_{\rm M}(\mv{b},t)\ast\left[\delta(t)-h_{\rm sol}(t) \right]\ast h_{\rm avg}(t)
\end{equation}
where $\delta(t)$ is the Dirac delta function, and $h_{\rm sol}$ is given by
\begin{equation}
h_{\rm sol}(t) = \begin{cases}
t_{\rm sol}^{-1} & \textrm{if } -t_{\rm sol}/2 < t < t_{\rm sol}/2 \\
0 & \textrm{otherwise}
\end{cases}
\end{equation}
The solution cadence $t_{\rm sol}$ can be far larger than the averaging interval $\Delta\tau_{\rm cell}$ and the visibilities in general will change over $t_{\rm sol}$. However calibration algorithms will take into account the rotation of baseline with time and hence automatically account for the corresponding change in the sky power spectrum $P_k(\mv{b},\Delta\mv{s}=\mv{v}\tau)$ from equation (\ref{ch4eqn:general_cov}). In order to follow the same steps as in Section \ref{ch4sec:fse} to evaluate the effects of self-calibration, we will however make the assumption that the \emph{geometry} of the projected baseline on the ionospheric screen does not change more than the Fresnel scale during the solution interval, such that equation (\ref{ch4eqn:general_cov}) can still be applied. We have chosen the Fresnel scale here since it is the natural coherence scale for our diffraction calculations. This assumption holds when $bt_{\rm sol}\lesssim 10^6$~m~s.  Under this assumption, by following the same steps as in section \ref{ch4sec:fse} and using the associative property of convolution, we can write
\begin{eqnarray}
\label{ch4eqn:solavg}
\sigma^2[\Delta V_{\rm C}(\mv{b})] &=& 4S^2_{\rm eff}\int {\rm d}^2\mv{q}\ips{\mv{q}} \sin^2(\upi\lambda h \mv{q}^2-\upi\mv{q}\bcdot\mv{b})\nonumber \\
&& \times \, \left[ \frac{\sin(\upi \Delta\tau_{\rm cell}\mv{q}\bcdot\mv{v})}{\upi \Delta\tau_{\rm cell} \mv{q}\bcdot\mv{v}}\right]^2 \nonumber \\
&& \times \, \left[1- \frac{\sin(\upi t_{\rm sol}\mv{q}\bcdot\mv{v})}{\upi t_{\rm sol} \mv{q}\bcdot\mv{v}}\right]^2
\end{eqnarray}
In general, the residual scintillation noise increases as one increases the solution cadence $t_{\rm sol}$. However in Fig. \ref{ch4fig:time_coherence} we saw that the averaging process suppresses power coming from small scale turbulence since the sinc$^2$ functions falls off for large $\mv{q}$. Equation (\ref{ch4eqn:solavg}) shows that calibration on the other hand suppresses power from large scale structure since the $(1-{\rm sinc})^2$ functions falls off for small values of $\mv{q}$. Hence one would intuitively choose $t_{\rm sol} \lesssim \Delta\tau_{\rm cell}$, such that the combined effects of self-calibration and averaging effectively mitigates scintillation noise\footnote{Not that we cannot choose an averaging interval $t_{\rm avg}\gtrsim \Delta\tau_{\rm cell}$ to avoid fringe decorrelation and loss of information.}. This fact is reflected in Fig. \ref{ch4fig:opt_tsol} where we plot the residual fractional scintillation noise as a function of the ratio between $t_{\rm sol}$ and $\Delta\tau_{\rm cell}$. The figure shows that for $t_{\rm sol}/\Delta\tau_{\rm cell} \sim 1$ self calibration mitigates about half of the scintillation noise rms. As one reduces $t_{\rm sol}$ further, the reduction in residual scintillation noise is logarithmic. For a $30$~meter aperture at $150$~MHz, $t_{\rm avg}=\Delta\tau_{\rm cell}$ decreases from about $34$~minutes to $1.7$~minutes as the baseline length increases from $100$~m to $2$~km. The corresponding values for a $5$~meter aperture are about $6$~minutes and $20$~s. The number of constraints available to any array within these time intervals will then determine the number of directions scintillation noise can be mitigated in. The required number of direction though depends on the angular coherence of scintillation noise.
\subsection{Direction dependent effects and angular coherence}
\label{ch4sec:dde}
\begin{figure*}
\centering
\includegraphics[width=\linewidth]{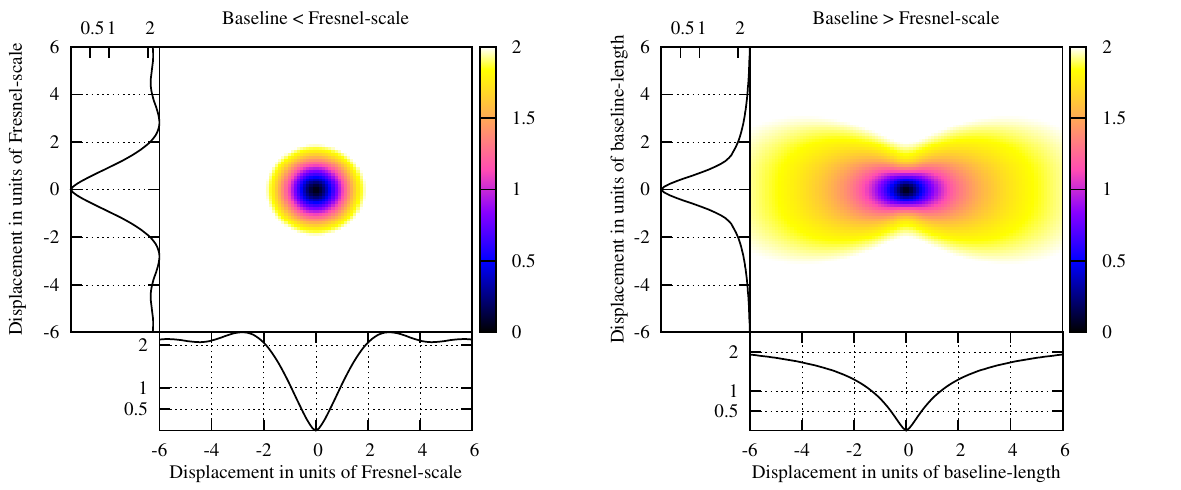}
\caption{Ratio of scintillation variance of a source after calibration transfer to its pre-calibration scintillation variance as a function of projected separation (on the ionospheric phase screen) between the source and the calibrator. Left and right panels are for baselines smaller and larger than the Fresnel scale respectively. Calibration transfer does more harm then good, if the projected separation exceeds $r_{\rm F}$ or $b$ respectively. The projected size of $r_{\rm F}$ varies from about $6$' to $3.5$' as frequency increases from $50$ to $150$~MHz.\label{ch4fig:calapply}}
\end{figure*}
In many scenarios, the EoR fields contain a bright point-like source at the field centre for precision calibration of instrumental and some ionospheric effects. Since the central source is typically very bright, one can assume that calibration solutions in the direction of that source can be obtained with a time cadence that is smaller than the typical scintillation decorrelation times-scales. It is then instructive to compute the effect of applying these solutions to the visibilities, which is equivalent to applying the solutions to all the sources in the field. To understand the effect of applying calibrated gains, we consider two sources with unit flux density separated by $\Delta\mv{l}=\mv{l}_1-\mv{l}_2$ on the sky. The measured visibility of the two sources is simply the super-position of their individual visibilities: 
\begin{equation}
V_{\rm M}(\mv{b})=g_1V_{1{\rm T}}+g_2V_{2{\rm T}}
\end{equation}
where $V_{1{\rm T}}$ and $V_{2{\rm T}}$ are the `true' uncorrupted visibilities of the two sources, and $g_1$ and $g_2$ are random variables that represent the presence of stochastic scintillation noise. Application of calibration gains obtained on the first source gives the corrected visibility of the second source as 
\begin{equation}
V_{2{\rm C}} = V_{2{\rm T}} \frac{g_2}{g_1}.
\end{equation}
We are interested in the variance of $g_2/g_1$ which is a ratio of two random variables. Obtaining the variance in closed form is difficult, but we can approximate the variance by Taylor expanding the quotient about the expected values of $g_1$ and $g_2$. The expression further simplifies since $g_1$ and $g_2$ have the same expected values and variances, and we get (proof in Appendix \ref{ch4sec:ch4appa})
\begin{equation}
\frac{\sigma^2(V_{2C})}{V^2_{2{\rm T}}} \approx \frac{2\sigma^2(g_1) - 2{\rm Cov}(g_1,g_2)}{\langle g_1\rangle^2}
\end{equation}
Using the generalised covariance expression from equation (\ref{ch4eqn:general_cov}), and noting that $\lr{g_1} \approx 1$ in the weak scattering regime, we get
\begin{eqnarray}
\label{ch4eqn:calapplyvar}
\frac{\sigma^2(V_{2C})}{V^2_{2{\rm T}}} &\approx & 8\int {\rm d}^2\mv{q}\ips{\mv{q}}\sin^2\left(\upi\lambda h \mv{q}^2-\upi\mv{b}\bcdot\mv{q} \right) \nonumber \\
&& \times \,\me{-\ii\upi h\mv{q}\bcdot\Delta\mv{l}+\ii\upi/2}\sin\left(\upi h \mv{q}\bcdot\Delta\mv{l}\right)
\end{eqnarray}
The integral vanishes as $\Delta\mv{l}$ approaches $\mv{0}$ as expected. Figure \ref{ch4fig:calapply} shows the numerically computed fractional variance $\sigma^2(V_{2C})/V^2_{2{\rm T}}$ for two limiting cases: short baselines ($b\lesssim r_{\rm F}$, left-panel) and long baselines ($b\gtrsim r_{\rm F}$, right-panel) as a function of projected separation of the two sources on the ionosphere $h\Delta\mv{l}$. As expected, the ratio increases with the projected separation $h\Delta\mv{l}$, and reaches a value of unity for $h\Delta\mv{l}_{\rm crit}=r_{\rm F}$ for $b\lesssim r_{\rm F}$ and $h\Delta\mv{l}_{\rm crit}=b$ for $b\gtrsim r_{\rm F}$. Hence if the angular separation between the source and the calibrator exceeds the critical value of $\Delta\mv{l}_{\rm crit}$ then calibration transfer increases scintillation noise rather than decrease it. For $b\lesssim r_{\rm F}$ where current arrays are most sensitive to the cosmological $21$-cm signal, this critical angular separation varies from about $6$~arcmin at $50$~MHz to about $3.5$~arcmin at $150$~MHz. Hence phase referencing using a bright calibrator source is effectively impossible with current arrays. While self calibration on a bright source may mitigate the effects of instrumental gains variations, it invariably leads to an increase in scintillation noise rms by a factor of $\sqrt{2}$. If instrumental gains do not fluctuate over timescales of $\Delta\tau_{\rm coh}$ (few seconds), then an optimal compromise would be to use the high time resolution self-calibration (on the bright calibrator) solutions to subtract the calibrator and its scintillation noise, but only apply a low-pass filtered gain solutions to the residual visibilities. 
\subsection{Calibratability for compact arrays}
\label{ch4sec:calibratability}
We now address the topic of calibratability of scintillation noise given an array configuration. Computing the `calibratability limit' for an arbitrary array configurations requires one to compute the coherence of scintillation noise between two baselines of arbitrary length and orientation. Obtaining such a covariance in closed form is involved and we do not attempt it here. Instead since scintillation noise on all baselines within a Fresnel scale ($r_{\rm F}=310$~m at $150$~MHz) is expected to be coherent, we will present arguments regarding calibratability of compact arrays wholly within a diameter of $r_{\rm F}$, while noting that most of the sensitivity to the $21$-cm power spectrum comes from short baselines\footnote{The arguments presented here may not be applicable to cases where a large number of `long' baselines are used to calibrate the short baselines within a compact core.} ($b\lesssim r_{\rm F}$).\\

For a primary aperture of diameter $d_{\rm prim}$, the field of view is given by 
\begin{equation}
\Omega_{\rm prim} = \frac{4\lambda^2}{\upi d^2_{\rm prim}}
\end{equation}
On short baselines, scintillation noise from two sources decorrelates, if their projected separation (on the ionosphere) exceeds $r_{\rm F}$. Hence, for optimal mitigation\footnote{We say optimal since scintillation noise cannot be completely mitigated in the presence of thermal noise} of scintillation noise, we have to obtain self-calibration solutions on each patch on the sky with solid angle
\begin{equation}
\Omega_{\rm crit}=\frac{\upi r_{\rm F}^2}{h^2}
\end{equation}
Hence the number of directions one has to solve for is given by 
\begin{equation}
\label{ch4eqn:kdir} 
N_{\rm dir}=16 r^2_{\rm F}/d^2_{\rm prim}.
\end{equation}
Given $N_{\rm prim}$ primary antenna elements, we have $N^2_{\rm prim}/2$ visibilities to solve for scintillation noise `gains' in the $N_{\rm dir}$ directions\footnote{Different visibility polarisations have the same scintillation noise realisation and do not give independent `constraints'.}. However, redundant visibilities in the Fourier plane do not contribute independent pieces of information. We therefore proceed by computing the maximum number of independent pieces of information available in the Fourier plane as follows. The autocorrelation function of a circular aperture of size $d_{\rm prim}$ has a half-power width of $d_{\rm prim}$, and a corresponding area of $\upi d_{\rm prim}^2/4$ in the Fourier plane. For an array wholly within a Fresnel scale $r_{\rm F}$, the total available area in the Fourier plane is $\upi r_{\rm F}^2$. Hence, the maximum number of available constraints is 
\begin{equation}
\label{ch4eqn:ncons}
N_{\rm cons} = \frac{1}{2}\frac{\upi r_{\rm F}^2}{\upi d_{\rm prim}^2/4} = \frac{2r_{\rm F}^2}{d_{\rm prim}^2},
\end{equation}
where the additional factor of $2$ accounts for dependent information contained in the conjugate visibilities. Clearly $N_{\rm dir} \geq N_{\rm cons}$, which implies that a compact array wholly within a Fresnel scale does not contain sufficient number of constraints to fully mitigate scintillation noise via self-calibration.\\

If the critical number of constraints are not available, a practical way forward is then to solve for ionospheric distortions in the direction of $N_{\rm bright}$ brightest sources. For the source counts of the form given in equation (\ref{ch4eqn:dnds_appflux}), the number of source with flux above a threshold $S_{\rm max}$ is given by
\begin{equation}
N(S>S_{\rm max}) = \int_{S_{\rm max}}C\nu^{-\beta}S^{-\alpha}B_{\rm eff} = \frac{C\nu^{-\beta}S_{\rm max}^{1-\alpha}B_{\rm eff}}{\alpha-1}
\end{equation}
If we solve in $N_{\rm bright}$ directions towards as many brightest sources, then we get a modified value for $S_{\rm max}$ of
\begin{equation}
S^{\rm cal}_{\rm max} = \left(\frac{(\alpha-1)N_{\rm bright}}{C\nu^{-\beta}B_{\rm eff}}\right)^{1/(1-\alpha)}
\end{equation}
Using the relationship between $S_{\rm max}$ and $S_{\rm eff}$ from equation (\ref{ch4eqn:seff}), the effective scintillating flux after direction dependent calibration is
\begin{equation}
S_{\rm eff}^{\rm cal} = S_{\rm eff}^{\rm cal}  N_{\rm bright}^{\frac{3-\alpha}{2(1-\alpha)}}.
\end{equation}
For minimally redundant arrays, we can assume $N_{\rm bright}=N_{\rm prim}^2/2$, whereas for maximally redundant arrays, we have from equation (\ref{ch4eqn:ncons}) $N_{\rm bright} = N_{\rm cons}$, which is typically less than $N_{\rm prim}^2/2$. For these two limiting cases, we can write the effective scintillating flux after direction dependent calibration as
\begin{eqnarray}
S_{\rm eff}^{\rm cal} & = S_{\rm eff}\left( \frac{N_{\rm prim}}{\sqrt{2}}\right)^{\frac{3-\alpha}{1-\alpha}}\,\,\, \textrm{ minimally redundant}\nonumber \\
S_{\rm eff}^{\rm cal} & = S_{\rm eff}\left( \frac{\sqrt{2}r_{\rm F}}{d_{\rm prim}}\right)^{\frac{3-\alpha}{1-\alpha}}\,\,\, \textrm{ maximally redundant}
\end{eqnarray}

The above for the typical value of $\alpha=2.5$ yields
\begin{eqnarray}
S_{\rm eff}^{\rm cal} &=& S_{\rm eff}\left( \frac{N_{\rm prim}}{\sqrt{2}}\right)^{-1/3}\,\,\, \textrm{ minimally redundant}\nonumber \\
S_{\rm eff}^{\rm cal} &=& S_{\rm eff}\left( \frac{\sqrt{2}r_{\rm F}}{d_{\rm prim}}\right)^{-1/3}\,\,\, \textrm{ maximally redundant}\nonumber \\
\end{eqnarray}
For the maximally redundant case, for $d_{\rm prim}=15$ and $30$~m, at $150$~MHz ($r_{\rm F}=310$~m), we get an effective scintillating flux reduction (due to calibration) of about $33$\% and $40$\%,, respectively. This is insufficient to bridge the gap between the corresponding thermal noise and scintillation noise curves of Fig. \ref{ch4fig:filled}. We therefore conclude that even the large number of constraints provided by the dense cores of arrays such as the SKA and HERA, may not aid in mitigating scintillation noise to a level at or below the thermal noise ($1$~MHz bandwidth). We however stress here again that scintillation noise in the weak scattering regime is a broadband phenomenon, hence arrays such as HERA and SKA that have a fully filled Fourier plane (on short baselines) within a snapshot may be able to remove this frequency-coherent scintillation noise along with smooth spectrum foregrounds in their foreground subtraction step. The same is not true for current arrays such as LOFAR and MWA who have a highly chromatic snapshot coverage in the Fourier plane. In the next section, we compute the scintillation noise power spectrum for such arrays.
%
%
%
%
%
%
\section{Scintillation noise power-spectrum}
\label{ch4sec:snps}
In this section we accumulate the results of the preceding sections to make scintillation noise predictions for LOFAR and MWA in the $2$-dimensional power spectrum space spanned by $k_{\perp}$ and $k_{||}$ which are the transverse and line of sight wavenumbers respectively. The $21$-cm power spectrum computation typically involves gridding the visibilities from the entire exposure into a regular $uv\nu$-grid, where $u$ and $v$ are the transverse Fourier modes measured by the interferometer, and $\nu$ corresponds to the line of sight distance\footnote{Since the $21$-cm signal is a spectral line, frequency corresponds to redshift with in turn corresponds to line of sight distance (within a small bandwidth)}. All visibilities that fall into a given $uv\nu$ cell are averaged to obtain $V_{\rm G}(u,v,\nu)$. A Fourier transform along the frequency dimensions then places the measurements in wavenumber co-ordinates in all $3$ dimensions: $\widetilde{V}_{\rm G}(u,v,\eta)$. The power spectrum is then computed as $\left|\widetilde{V}_{\rm G}(u,v,\eta) \right|^2$. Many of the instrumental and ionospheric effects are best represented in a $2$-dimensional power spectrum that is obtained by averaging $\left|\widetilde{V}_{\rm G}(u,v,\eta) \right|^2$ within annuli in the $uv$ plane at each $\eta$. To forecast the ratio between scintillation and thermal noise, we use equation (\ref{ch4eqn:scthnoise}), with the appropriate scaling between $\eta$ and $k_{||}$, and between $b$ and $k_{\perp}$ at each redshift.\\

We assume a synthesis bandwidth of $10$~MHz for each redshift bin, since we do not expect significant power spectrum evolution within the associated redshift range. We present results for `bad', `moderate' and `good' ionospheric conditions. Based on statistics of diffractive scale measurements (with LOFAR data on 3C196) accumulated over several nights \citep{mevius}, we have chosen diffractive scale values at $150$~MHz of $r_{\rm diff}=5,10,20$~km to represent bad, moderate, and good nights respectively. We caution the reader that it is not uncommon to find epochs where $r_{\rm diff}\ll 5$~km, but such data typically show strong diffractive scintillation and must not be used in EoR analysis.\\ 

\begin{figure*}
\centering
\includegraphics[width=0.85\linewidth]{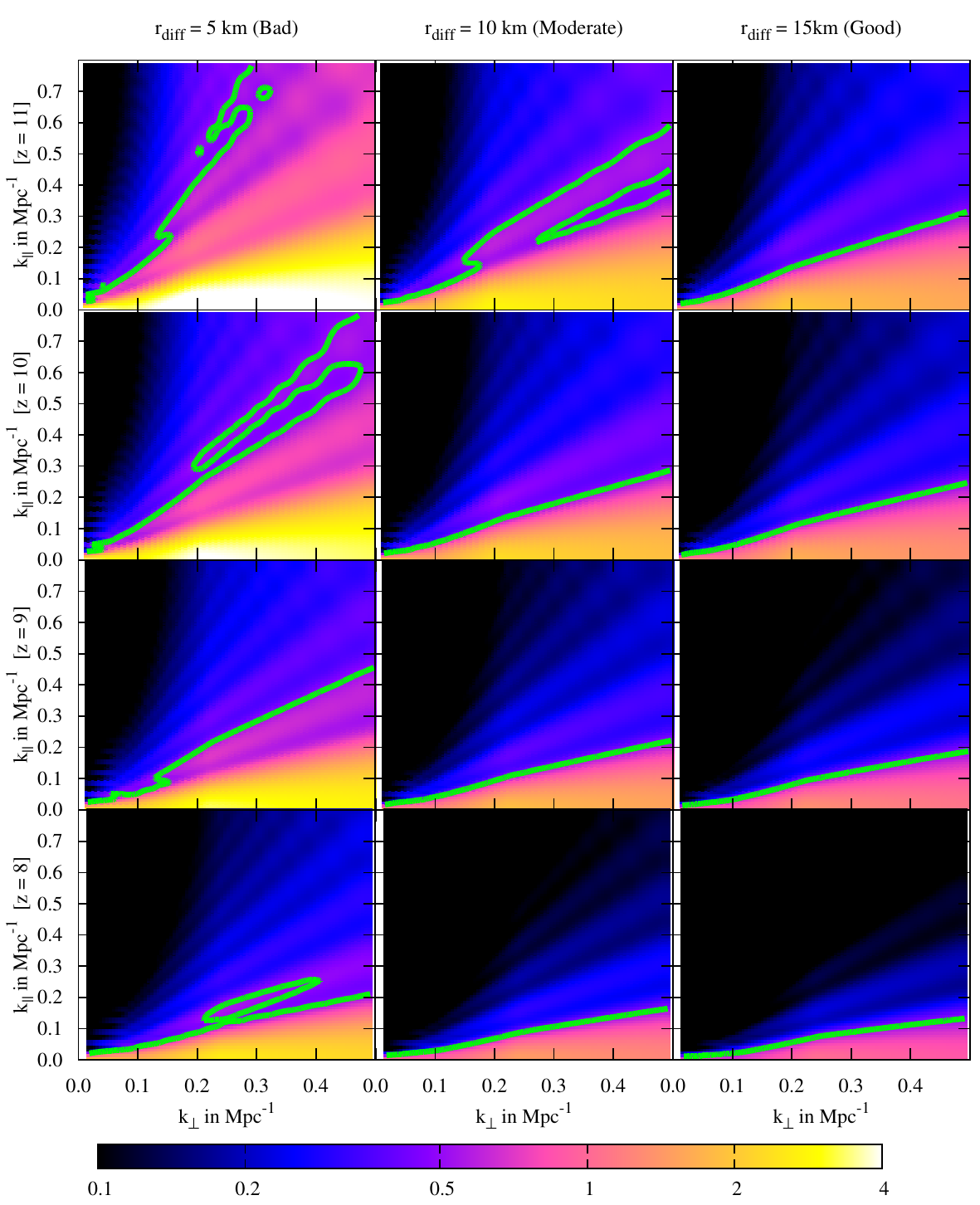}
\caption{Expected scintillation noise to thermal noise ratio cast in cosmological line of sight ($k_{||}$) and transverse ($k_{\perp}$) wavenumber axes. The parameters assumed here are summarised in table \ref{ch4tab:lofar_ncp} and are representative of LOFAR observations of the NCP field. The contour line traces a ratio of $1/2$.\label{ch4fig:kpkp_lofar}}
\end{figure*}

\begin{figure*}
\centering
\includegraphics[width=0.85\linewidth]{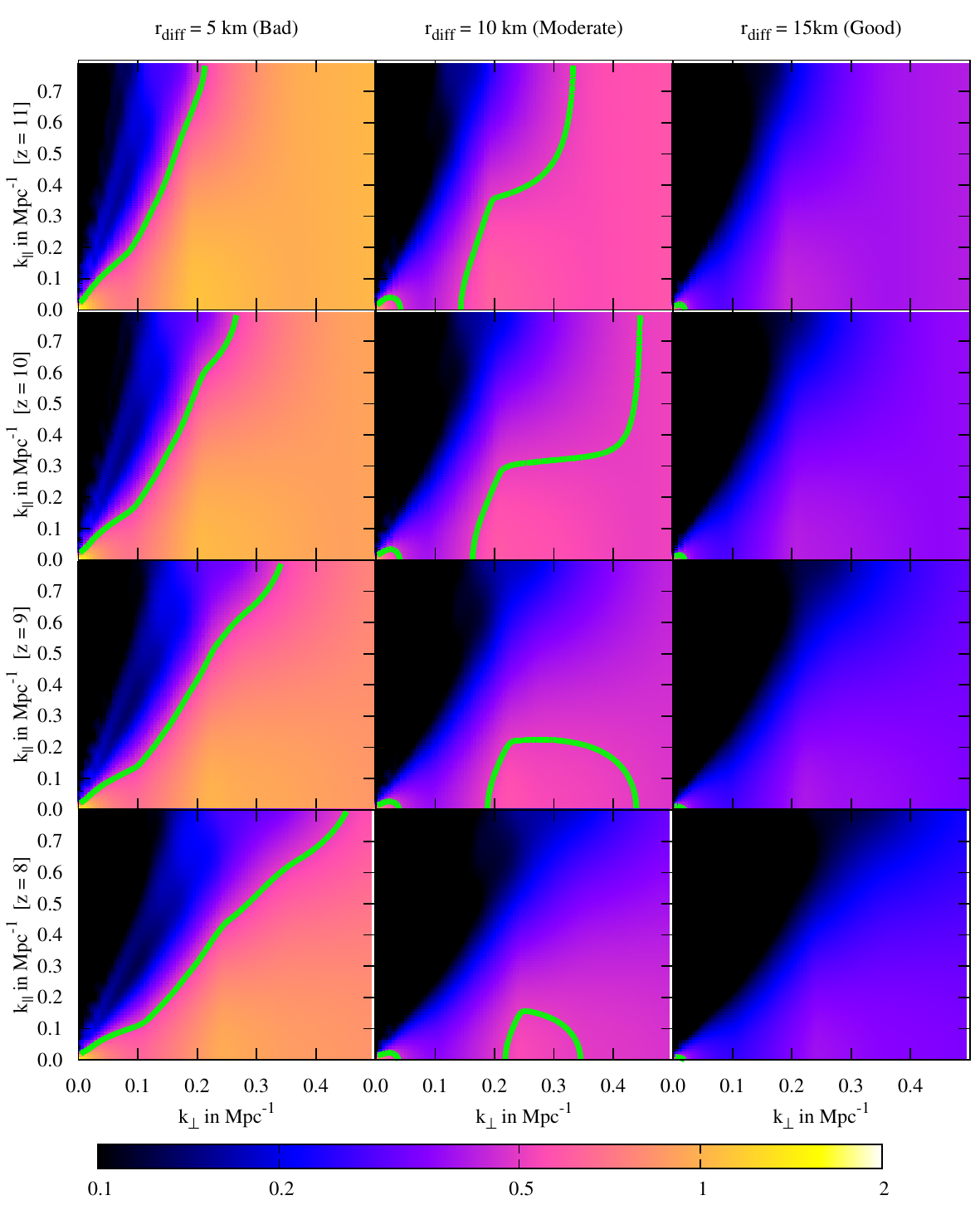}
\caption{Same as Fig. \ref{ch4fig:kpkp_lofar} but for parameters (summarised in table \ref{ch4tab:mwa_zenith}) representative of MWA observations at zenith.\label{ch4fig:kpkp_mwa}}
\end{figure*}

Figure \ref{ch4fig:kpkp_lofar} shows the ratio between the scintillation noise and thermal noise:
\[
\sigma_{\rm sc}(k_{\perp},k_{||})/\sigma_{\rm th}(k_{\perp},k_{||})
\]
for different redshifts (rows) and ionospheric conditions. We have assumed parameters (summarised in table \ref{ch4tab:lofar_ncp}) representative of LOFAR observations of the North Celestial Pole. Figure \ref{ch4fig:kpkp_mwa} shows the same ratio but for parameters (summarised in table \ref{ch4tab:mwa_zenith}) representative of MWA observations with a zenith pointing. In both cases, we have not assumed any mitigation of scintillation noise by direction dependent self-calibration. Hence the plots represent the condition $t_{\rm sol}\gtrsim \Delta\tau_{\rm coh}$ at the longest baselines that goes into the $21$-cm power spectrum analysis, and as such may be considered the worst case scenario, assuming that calibration is not affected by scintillation noise on these baselines. The iso-contour line traces a ratio of $1/2$ at which we have to integrate for $25$\% longer to achieve the same power spectrum sensitivity as previously thought (thermal noise alone). Hence the region to the right and below the contour is expected to have a considerable impact on the sensitivity to the $21$-cm power spectrum signal. We again stress here that the delay-domain sidelobes can be largely mitigated with a suitable window function (see Fig. \ref{ch4fig:nf_wf_eta}). As expected, scintillation noise follows the well understood `wedge' like structure in the $k_{\perp},k_{||}$ space, and is about the same order of magnitude (within the wedge) as thermal noise. 
\begin{table}
\centering
\caption{Parameters used to generate Fig. \ref{ch4fig:kpkp_lofar} representative of LOFAR observations of the NCP field \label{ch4tab:lofar_ncp}}
\begin{tabular}{ll}
Parameter & Value \\ \hline
$\theta$ & $38$~deg\\
$\theta$ scaling of $\sigma_{\rm sc}$ & $\sec^{11/12}\theta$\\
SEFD & 3807 Jy (freq. independent)\\
$S_{\rm max}$ & $9.5$~Jy (freq. independent)\\
$S_{\rm eff}$ & equation (\ref{ch4eqn:seff}) with above $S_{\rm max}$\\
$d_{\rm prim}$ & 30~m\\
$v$ & 500~km~hr$^{-1}$\\
$\sigma_{\rm sc}$ & equation (\ref{ch4eqn:fullsigmasc})\\
$\sigma_{\rm th}$ & equation (\ref{ch4eqn:fullsigmath})\\
$t_{\rm sol}$ & $\gtrsim 2b_{\rm max}/v$ ($=20$~sec)\\ \hline
\end{tabular}
\end{table}
\begin{table}
\centering
\caption{Parameters used to generate Fig. \ref{ch4fig:kpkp_mwa} representative of MWA observations at zenith \label{ch4tab:mwa_zenith}}
\begin{tabular}{ll}
Parameter & Value \\ \hline
$\theta$ & $0$~deg\\
SEFD & equation (\ref{ch4eqn:sefdvals})\\
$S_{\rm eff}$ & equation (\ref{ch4eqn:seffvals})\\
$d_{\rm prim}$ & 5~m\\
$v$ & 500~km~hr$^{-1}$\\
$\sigma_{\rm sc}$ & equation (\ref{ch4eqn:fullsigmasc})\\
$\sigma_{\rm th}$ & equation (\ref{ch4eqn:fullsigmath})\\
$t_{\rm sol}$ & $\gtrsim 2b_{\rm max}/v$ ($=20$~sec)\\ \hline
\end{tabular}
\end{table}
%
%
%
%
%
%
%
\section{Main Results}
\label{ch4sec:mainresults}
In this paper, we have used the analytical results from \citet{speckle} to compute the scintillation noise bias in $21$-cm power spectrum measurements. In the process, we have discussed the implications of various data processing steps such as averaging, calibration, gridding (during Fourier synthesis) for scintillation noise. We have arrived at the following conclusions.
\begin{enumerate}
\item Scintillation noise from a `sea' of point sources that follow a certain source-counts law (equation (\ref{ch4eqn:dsc})) is equal to scintillation noise of a single source of flux $S_{\rm eff}$ which is the rms apparent-flux of all sources in the sky, since both have the same power spectrum.  The value of $S_{\rm eff}$ is mostly influenced by the high-flux end of the source distribution and is related to the flux-density of the brightest source contributing to the scintillation noise through equation (\ref{ch4eqn:seff}).
\item Off zenith viewing geometry leads to an increase in scintillation noise variance by a factor of $\sec^{11/6}\theta$ due to (i) increased path-length through the ionosphere, and (ii) increase in distance to the effective ionospheric phase-screen leading to an increase in the Fresnel-length (Fig. \ref{ch4fig:za_dependence} and table \ref{ch4tab:za_scaling}). However, for zenith viewing telescopes, the analytical expressions derived in \citet{speckle} are valid approximations even for arbitrarily large fields of view.
\item The monochromatic scintillation and thermal noise evaluated within a scintillation decorrelation timescale, and fringe decorrelation bandwidth is given by equation (\ref{ch4eqn:scthnoise}). After applying a delay transform (Fourier transform along frequency) to the gridded visibility data, the scintillation and thermal noise contribution from a single baseline to a $uv\eta$-cell is given by equations (\ref{ch4eqn:fullsigmasc}) and (\ref{ch4eqn:fullsigmath}). 
\item Though scintillation noise is a broadband phenomena, its sampling in the Fourier plane is generally not so. This is because snapshot $uv$-coverage of minimally redundant arrays such as LOFAR and MWA is poorly filled leading to a spatial sampling function in the $uv$-plane that `stretches' with frequency. This leads to decorrelation of measured scintillation noise (Fig. \ref{ch4fig:freq_coherence}) over the fringe decorrelation bandwidth ($\Delta\nu_{\rm cell}$ from equation (\ref{ch4eqn:nucell})). Hence spectral coherence properties of scintillation noise also follow those of sidelobe noise leading to a well established `wedge' structure in the $2$-dimensional (cylindrical) power spectrum \citep[][and Fig. \ref{ch4fig:nf_wf_eta}]{vedantham2012}. It is important to realise that though the $uv$-plane may be completely filled at all frequencies after Earth rotation synthesis, since scintillation is a time variable phenomena, the relevant $uv$-coverage to consider here is the \emph{snapshot} $uv$-coverage. Scintillation noise leaks above the wedge by virtue of the sidelobes of the delay transform point spread function\footnote{Which is also the Fourier transform of the bandpass window function}, and can be mitigated in this region using a suitable window function in the delay transform \citep{vedantham2012}.  
\item Although filled apertures such as HERA and SKA-LOW (core only) will be scintillation noise dominated (Fig. \ref{ch4fig:filled}), because this scintillation noise is correlated on all baselines of a compact array wholly within a Fresnel scale ($r_{\rm F}=310$~m at $150$~MHz), the resulting scintillation noise manifests as an uncertainty in the flux emanating from every patch of the sky of solid angle $\upi r_{\rm F}^2/h^2$ ($h$ is distance to the ionosphere). However such arrays have a completely filled \emph{snapshot} $uv$-coverage, and the scintillation noise they measure will have large frequency coherence (weak scintillation), which will probably enable it to be subtracted away along with smooth-spectrum astrophysical foregrounds.
\item During Earth rotation synthesis, visibilities are averaged over timescales of $\Delta\tau_{\rm cell}$ (equation (\ref{ch4eqn:taucell})). While this time averaging leads to suppression of scintillation from small-scale turbulence (Fig. \ref{ch4fig:time_coherence}), self-calibration leads to suppression of contribution from large-scale wavemodes. If the ratio of solution cadence to averaging interval $t_{\rm sol}/\Delta\tau_{\rm cell}= 1$ then scintillation noise is mitigated by about $50$\% in rms (Fig. \ref{ch4fig:opt_tsol}). Decreasing solution cadence further logarithmically reduces residual scintillation noise.
\item Efficiency of scintillation noise mitigation using self-calibration solution transfer from a calibrator to a target source depends on the projected separation between the sources on the phase-screen $\Delta\mv{s}=h\Delta\mv{l}$ where $h$ is the distance to the phase screen and $\Delta\mv{l}$ is the angular separation of the two sources (Fig. \ref{ch4fig:calapply}). For $b\lesssim r_{\rm F}$ calibration transfer does more harm then good if $|\Delta\mv{s}|\gtrsim r_{\rm F}$. For $b\gtrsim r_{\rm F}$ the same is true for $|\Delta\mv{s}|\gtrsim b$.
\item The above result and the angular coherence of scintillation noise set a lower limit of $16 r_{\rm F}^2/d_{\rm prim}^2$ on number of direction one has to solve for in self-calibration to mitigate scintillation noise to at or below the thermal noise in compact arrays wholly within a Fresnel scale (see equation (\ref{ch4eqn:kdir})). The number of independent visibility constraints available in such an array is insufficient to solve in as many directions. 
\item Due to the above lack of constraints, one can solve for the scintillation noise from the brightest $N_{\rm bright}$ sources, where $N_{\rm bright}$ scales as $N^2_{\rm prim}/2$ and is bounded by a maximum value of $N_{\rm bright} = 2r_{\rm F}^2/d_{\rm prim}^2$. Such a solution will reduce the effective scintillating flux by a factor of $N_{\rm bright}^{-1/6}$. Regardless, as stated earlier, the measured scintillation noise for arrays with a fully-filled snapshot $uv$-coverage such as SKA (core) and HERA is broadband, which may allow it to be subtracted along with smooth-spectrum foreground emission. 
\item Figures \ref{ch4fig:kpkp_lofar} and \ref{ch4fig:kpkp_mwa} shows the predicted ratio of scintillation to thermal noise at different redshifts for differing ionospheric conditions. We have chosen the parameters for the two figures (summarised in table \ref{ch4tab:lofar_ncp} and \ref{ch4tab:mwa_zenith}) to represent the case of LOFAR observations of the NCP and MWA observations at zenith respectively. The region to the left and bottom of the iso-contour lines (drawn at a ratio of $1/2$) will require significant increases ($>25$\%) in integration time to attain the same sensitivity as previously thought (based on thermal noise alone). \\
\end{enumerate}
\section{Conclusions and future work}
\label{ch4sec:concl}
In this paper many new results have been derived, summarised in the previous section. Below we list a number of higher-level conclusions and shortly discuss directions for future work.
\begin{enumerate}
\item Ionospheric scintillation noise in the cylindrical power spectrum space for current arrays (e.g. LOFAR and MWA) is largely confined below the `wedge' \citep{vedantham2012} for the case of weak scattering ($r_{\rm diff}\gtrsim b_{\rm max}$), and as such does not pose a fundamental limitation to current $21$-cm power spectrum efforts.
\item Assuming `primary calibration' removes scintillation noise from very bright sources (those that present a signal to noise ($1$~MHz, $2$~s interval) ratio of S/N~$\ga 5$ per visibility), scintillation noise within the wedge, if unmitigated, will require at least $25$\% (and in some regimes $100$\%) more integration time to achieve the same power spectrum sensitivity as previously thought.
\item Mitigating ionospheric effects via phase referencing using a bright calibrator is not an option for high redshift $21$-cm experiments, due to a small coherence angle of ionospheric phase fluctuations on baselines where the signal is expected to be the strongest ($\lesssim$~few km). 
\item The fully filled cores of HERA and SKA will be scintillation noise dominated on all baselines. Since the cores are mostly confined to a size $\lesssim r_{\rm F}$, all ionospheric effects are coherent in space and frequency, although incoherent between most sources. This frequency coherence allows scintillation noise to be largely subtracted together with the smooth foregrounds in the frequency direction. 
\item Computing the efficacy of direction-dependent calibration in mitigating scintillation noise for an arbitrary array configuration will be instrumental in evaluating the utility of longer baselines in mitigating ionospheric effects on shorter baselines which contain the bulk of the cosmic $21$-cm signal. This forms an important part of current research efforts.
\item Antenna based calibration solutions may not be the most efficient technique to mitigate scintillation noise on short baselines ($b\lesssim r_{\rm F}$) since they experience direction-incoherent but baselines-coherent scintillation noise. Our future efforts involve developing and testing such a source-dependent, baseline-independent calibration algorithm.
\end{enumerate}
In this paper, we have worked out the coherence function of ionospheric phase fluctuations in the temporal, frequency, baseline and directional domain. Our main conclusion is that, although ionospheric phase errors can add an additional bias to the residual power-spectrum in 21-cm (or other) observations, under reasonable observing conditions they will not pose a `show-stopper'. Its impact however needs careful study, since the true impact of the ionosphere might not come via direct speckle or scintillation noise from the sky, but from gain solutions derived from an imperfect sky model coupled to ionospheric phase errors. This remains a topic for future research.

\section*{Acknowledgements}
The authors thank Dr. Maaijke Mevius and Professor A.G. de Bruyn for many interesting discussions. The authors also acknowledge the financial support from the European Research Council under ERC-Starting Grant FIRSTLIGHT - 258942.

\bibliographystyle{mn2e}
\bibliography{mombib.bib}

\section*{Appendix: Variance of a ratio}
\label{ch4sec:ch4appa}
We are interested in computing the variance $\sigma^2(g_1/g2)$ where $g_1$ and $g_2$ are random variables with means $\mu_1$ and $\mu_2$ and variances $\sigma^2(g_1)$ and $\sigma^2(g_2)$ respectively. Obtaining the variance of the ratio in closed form is difficult, but we can obtain a good approximation by Taylor expanding the ratio about $\mu_1/\mu_2$ as
\begin{equation}
\frac{g_1}{g_2} \approx \frac{\mu_1}{\mu_2} + \frac{g_1-\mu_1}{\mu_2}-\frac{\mu_1}{\mu_2^2}(g_2-\mu_2).
\end{equation} 
The variance of the ratio is thus
\begin{equation}
\sigma^2\left(\frac{g_1}{g_2}\right) \approx \sigma^2\left(\frac{\mu_1}{\mu_2}+ \frac{g_1}{\mu_2}+\frac{\mu_1g_2}{\mu_2^2}\right)
\end{equation}
which on using $\sigma^2(a\pm b) = \sigma^2(a) + \sigma^2(b) \pm 2 {\rm Cov}(a,b)$ yields
\begin{equation}
\sigma^2\left(\frac{g_1}{g_2} \right) \approx \frac{1}{\mu_2^2}\left( \sigma^2(g_1) + \frac{\mu_1^2}{\mu_2^2}\sigma^2(g_2) -2 \frac{\mu_1^2}{\mu_2^2}{\rm Cov}(g_1,g_2)\right).
\end{equation}
For the case of scintillation from a spatially stationary ionosphere, we have $\mu_1=\mu_2=\lr{g_1}$ say, and $\sigma^2(g_1) = \sigma^2(g_2)$. Using this we get
\begin{equation}
\sigma^2\left(\frac{g_1}{g_2} \right) \approx \frac{2\sigma^2(g_1)-2{\rm Cov}(g_1,g_2)}{\lr{g_1}^2}
\end{equation}

\begin{table*}
\centering
\caption{A glossary of terms and their meaning}
\begin{tabular}{lp{10cm}}
\hline
Term & Meaning \\ \hline
$\nu$ & Electromagnetic wave frequency\\
$\lambda $ & Electromagnetic wavelength\\
$\ips{\mv{q}}$ & Ionospheric phase power spectrum where $\mv{q}$ is the wavenumber vector.\\
$\phi_0^2$ & Ionospheric phase variance \\
$k_{\rm o}$ & Outer-scale (of turbulence) wavenumber\\
$r_{\rm diff}$ & Ionospheric diffractive scale\\
$\mv{v}$ & Ionospheric bulk-velocity \\
$ S_{\rm eff}$ & Effective scintillating flux\\
$\alpha$ & Spectral index with which differential source counts scale with flux density\\
$\beta$ & Spectral index with which differential source counts scale with frequency\\
$\gamma$ & Spectral index with which sky brightness temperature scales with frequency\\
$ r_{\rm F}$ & Fresnel-scale\\
$\theta$ & Zenith angle\\
$\sigma^2_{\rm rf}[\mv{b}]$ & Scintillation variance due to a 1~Jy source at the phase centre.\\
& Also called fractional scintillation variance.\\
$\mv{l}$ & 2D direction cosine vector\\
$ h(\theta)$ & Distance to ionospheric phase screen \\
$\mv{b}$ & 2D baseline vector\\
$B(\nu,\mv{l})$ & Primary beam of interferometer element\\
$B_{\rm eff}(\nu)$ & Effective beam for scintillation calculations\\
$\Delta u_{\rm cell},\Delta v_{\rm cell}$ & $uv$-plane cell size\\
$\Delta\tau_{\rm cell}(b)$ & Time spent by a baseline (length $=b$) in a $uv$-cell.\\
$d_{\rm prim}$ & Aperture diameter of primary antenna\\
$\Delta \nu_{\rm cell}(b)$ & Frequency interval spend by a baseline (length $=b$) in a $uv$-cell\\
SEFD & System equivalent flux density\\
$\Delta\tau_{\rm coh}(b)$  & Scintillation noise coherence time-scale\\
$t_{\rm avg}$ & Visibility time-averaging interval\\
$\Delta\nu_{\rm ch}$ & Visibility frequency-averaging interval\\
$t_{\rm sol}$ & Self-calibration solution-time interval\\
$d_{\rm core}$ & Array diameter for a maximally redundant array\\
$N_{\rm prim}$ & Number of interferometer elements in a maximally redundant array\\
$\sigma_{\rm th}$ & Thermal noise standard deviation\\
$\sigma_{\rm sc}$ & Scintillation noise standard deviation\\
$V_{\rm T}$ & Visibility in the absence of scintillation\\
$V_{\rm M}$ & Visibility corrupted by scintillation\\
$V_{\rm C}$ & Visibility after application of self-calibration solutions\\ \hline
\end{tabular}
\end{table*}

\end{document}